# Mid-Infrared Bi-directional Reflectance Spectroscopy of Impact Melt Glasses and Tektites

**Corresponding Author:** Andreas Morlok, Institut für Planetologie, Wilhelm-Klemm-Str. 10, 48149 Münster, Germany. Email: morlokan@uni-muenster.de, Tel. +49-251-83-39069

Aleksandra Stojic, Institut für Planetologie, Wilhelm-Klemm-Str. 10, 48149 Münster, Germany. Email: a.stojic@uni-muenster.de

Iris Weber, Institut für Planetologie, Wilhelm-Klemm-Str. 10, 48149 Münster, Germany. Email: sonderm@uni-muenster.de

Harald Hiesinger, Institut für Planetologie, Wilhelm-Klemm-Str. 10, 48149 Münster, Germany. Email: hiesinger@uni-muenster.de

Michael Zanetti, University of Western Ontario, 1151 Richmond St, London, Ontario, Canada N6A 3K7. Email: mzanett3@uwo.ca

Joern Helbert, Institute for Planetary Research, DLR, Rutherfordstrasse 2, 12489 Berlin, Germany, Email: joern.helbert@dlr.de

Keywords: Spectroscopy; Impact processes; Instrumentation; Infrared Observations; Mercury

**Abstract**

We have analyzed 14 impact melt glass samples, covering the compositional range from highly felsic to mafic/basaltic, as part of our effort to provide mid-infrared spectra (7-14 µm) for MERTIS (Mercury Radiometer and Thermal Infrared Spectrometer), an instrument onboard of the ESA/JAXA BepiColombo mission.

Since Mercury was exposed to many impacts in its history, and impact glasses are also common on other bodies, powders of tektites (Irghizite, Libyan Desert Glass, Moldavite, Muong Nong, Thailandite) and impact glasses (from the Dellen, El'gygytgyn, Lonar, Mien, Mistastin, and Popigai impact structures) were analyzed in four size fractions of (0-25, 25-63, 93-125 and 125-250 µm) from 2.5-19 µm in bi-directional reflectance. The characteristic Christiansen Feature (CF) is identified between 7.3 µm (Libyan Desert Glass) and 8.2 µm (Dellen). Most samples show mid-infrared spectra typical of highly amorphous material, dominated by a strong Reststrahlen Band (RB) between 8.9 µm (Libyan Desert Glass) and 10.3 µm (Dellen). Even substantial amounts of mineral fragments hardly affect this general band shape.

Comparisons of the $SiO_2$ content representing the felsic/mafic composition of the samples with the CF shows felsic/intermediate glass and tektites forming a big group, and comparatively mafic samples a second one. An additional sign of a highly amorphous state is the lack of features at wavelengths longer than ~15 µm. The tektites and two impact glasses, Irghizite and El'gygytgyn respectively, have much weaker water features than most of the other impact glasses.

For the application in remote sensing, spectral features have to be correlated with compositional characteristics of the materials. The dominating RB in the 7-14 µm range correlates well with the $SiO_2$ content, the Christiansen Feature shows similar dependencies. To distinguish between glass and crystalline phases of the same chemical composition, a comparison between CF the SCFM index ($SiO_2/(SiO_2+CaO+FeO+MgO)$) (Walter and Salisbury, 1989) is useful, if chemical compositional data are also available.

# 1. Introduction

This study focuses on mid-infrared reflectance spectra of melt glasses from impact events for the application in planetary remote sensing. The obtained spectra will be part of a database for the ESA/JAXA BepiColombo mission to enter the Hermean orbit in 2024 (Maturilli et al., 2008, Benkhoff et al., 2010), and general remote sensing applications. The ESA/JAXA BepiColombo mission includes a mid-infrared spectrometer (MERTIS-Mercury Radiometer and Thermal Infrared Spectrometer) that allows for mapping the surface of Mercury in the 7-14 µm range, with a spatial resolution of ~500 m (Helbert et al., 2009; Benkhoff et al., 2010; Hiesinger et al., 2010). Laboratory FTIR (Fourier Transformed Infrared) spectra of minerals and rocks, but also of synthetic analogs, have to be collected to be compared to those from MERTIS for the interpretation of the remote sensing data from the surface of Mercury (For a discussion of the comparison of bi-directional reflectance and emission data see below, 2.6 Bi-directional reflectance FTIR-Analyses).

The surfaces of terrestrial planets and their moons are shaped and modified by impact events throughout their lifetimes (Hörz and Cintala, 1997). Thus, the investigation of how these processes affect the spectral properties of the rocks, which are modified or produced during the impact event is important for the interpretation of infrared data from planetary bodies. In particular, higher shock ranges result in amorphous phases produced in solid state transformation (such as maskelynite), or melt glass from complete melting of the target material (e.g. Stöffler, 1966; Ostertag, 1983; Koeberl, 1986; Dressler and Reimold, 2001; Osinski et al., 2008; Wünnemann et al., 2008; Osinski and Pierrazo, 2012; Jaret et al., 2015a; Pickersgill et al., 2015).

The aim of this study is to provide mid-infrared spectra in the 7-14 µm range for bulk impact melt rocks, as well as separated glasses for four size fractions (0-25, 25-63, 63-125, 125-250 µm). Motivation for studying several size fractions is that a high porosity and large grain size variations are characteristic properties of surface regolith material, which affects the spectral properties of the materials and thus have to be taken into account in the interpretation of remote sensing data.

Variation in grain size cause the intensity of the characteristic Reststrahlen Bands (RB), fundamental mode absorption features in the 7-14 µm region, to loose spectral contrast with decreasing grain size, so the intensity of the bands gets weaker (e.g., Salisbury and Eastes, 1985; Salisbury and Wald, 1992; Mustard and Hayes, 1997; Ruff and Christensen, 2002). As a consequence, the identification of phases based on band positions my get difficult. Earlier studies in the visible and near-infrared indicate a strong influence of grains smaller 30 µm in sizes for the surface regolith on Mercury (Sprague et al., 2007), so the RB can expected to be weak in the remote sensing data of this planet. Furthermore, a characteristic additional spectral feature, the Transparency Feature (TF) appears around 11 – 13 µm in the smallest grain size fractions as a result of increased volume scattering (e.g. Salisbury, 1993). Such features may have been observed in earlier, ground based infrared observation of Mercury. Thus laboratory spectra showing this diagnostic band are also of high interest (e.g. Cooper et al., 2001; Sprague et al., 2007).

So a series of analyses of distinct size fractions is needed to distinguish between genuine effects and effects induced by mixed grain size fractions in the natural regolith. However, since regolith is also an intimate mixture of many phases, spectral unmixing modelling will also have to take many other grain-size fractions of other potential mineral phases into account, which will complicate the discussion of the results.

The present work is a follow-up on an earlier study, where we analyzed Suevite impact rocks from the Nördlinger Ries. The former study shows a high degree of amorphization among samples of the Suevite impact rock that contains material in all stages of shock metamorphism (Morlok et al., 2016). Here we focus mostly on impact glasses formed from quenched impacts melts as representatives of the highest shock stages. In shock metamorphism, mineral phases start to change above 2 GPa. Planar deformation features appear between 8 and 25 GPa. In addition, microscopic changes in the crystal structure of quartz and feldspar arise. At pressures between 25 -40 GPa, minerals turn into diaplectic glasses (e.g. maskelynite)-in solid-state transformation, forming of and amorphous material. Onset of melting of phases starts with feldspar at pressures from ~35 to ~45 GPa, over 60 GPa rocks completely melt. Vaporization follows at pressures above 100 GPa (e.g., Stöffler, 1966, 1971, 1984; Chao, 1967; von Engelhardt and Stöffler, 1968; Stöffler and Langenhorst, 1994; French, 1998; Johnson, 2012). However, these pressures are only approximate to provide a general picture. Shock pressures and therefore their effects on minerals are often heterogeneously distributed in the rocks (e.g. Hanss et al., 1978; Johnson et al., 2002). Furthermore, they are based on crystalline, non-porous rocks. Different abundances of mineral species like quartz and feldspar could affect the exact pressure range since they show differences in shock transformation (Ostertag, 1983; Stöffler and Langenhorst, 1984). Petrology (Hörz and Cintala, 1997) but also structural features like grains size or porosity might play a role (French, 1998; Wünnemeann et al., 2008). Also, variations in mineral chemical composition are also important. For example, Ca-rich plagioclase transforms into maskelynite at lower pressures than K- and Na-rich feldspars (e.g. Ostertag, 1983, Johnson and Hörz 2003, Johnson, 2012).

In order to obtain a more comprehensive understanding of the spectral properties of highly shocked rocks, we present new mid-infrared spectra of impact glasses originating from a series of terrestrial impact craters. These glasses are found in the impactites in crater structures, but are also derived from distant ejecta, the so-called tektites (Dressler and Reimold, 2001).

We selected impact glasses and tektites from a series of terrestrial craters (for all details see Table 1) in the age range from 0.6 Ma (Lonar crater, Jourdan et al., 2011) to 119 Ma (Mien impact structure, Bottomley et al., 1978). The size of the respective impact craters or structures (if known) ranges from 1.8 km for Lonar crater (Wright et al., 2011) to about 100 km for the 36 Ma old Popigai impact structure (Bottomley et al., 1997; Kettrup et al., 2003). The Canadian Mistastin Lake impact structure is represented in our study by two melt glass samples (Grieve, 1975; Marion et al., 2011; Pickersgill et al., 2015).

The composition of the basement rocks, which are the assumed source material of melts, is varied, but in most craters dominated by granodioritic and gneissic basement rocks. The El'gygytgyn impact structure, is an exception with the basement being dominated by rhyolitic ignimbrites (Layer 2000; Wittmann et al., 2013; Raschke et al., 2014). The Lonar crater formed in basaltic basement material (Wright et al., 2011). Irghizite from the Zhamanshin impact structure (Deino et al., 1990) formed in a very diverse target material, consisting of schist, phyllite, clays and sandstones (Deino et al., 1990; Magna et al., 2011; Ackerman et al., 2015).

A special type of impact glass is called tektite, which is usually very pure, homogenous $SiO_2$-rich glass with low abundances of inclusions and typically particles display an aerodynamic shape (e.g., Koeberl, 1986; French, 1998; Dressler and Reimold, 2001). In our study they are represented by samples from two strewnfields. The Central European strewnfield is represented by Moldavites. Thailandites and Muong Nong tektites are from the Australasian strewnfield. Further strewnfields not covered in this study are the Ivory Coast and North American ones (French, 1998).

In the case of the Australasian tektites (0.77 Ma, Izett and Obradovich, 1992) the large required source crater (up to 116 km diameter) is unknown (Lee and Wei, 2000; Son and Koeberl, 2005). Boron isotope studies point towards marine or river sediments as source material (Chaussidon and Koeberl, 1995). The probable source materials of Moldavites were likely surface sediments of the Ries impact site, mainly composed of sands and clays (e.g., von Engelhardt et al., 2005).

Impact glass from the Nördlinger Ries crater is represented by two samples. Otting Glasbombe is glass separated from Suevite originating in the Otting site in the Ries, while the Polsingen sample is from the Red Suevite, basically a coherent melt rock (Stöffler et al., 2013; Morlok et al, 2016).

Similarly, the source crater for the Libyan Desert glasses is also unknown, but their origin from an impact into a sand/sandstone type material is strongly indicated by the occurrence of christobalite, meteoritic components (e.g., chondritic ratios of Co, Cr, Fe, Ni, and Ir) and high-temperature phases like baddeleyite ($ZrO_2$) (e.g., Barnes and Underwood, 1976; Storzer and Koeberl, 1991; Rocchia et al., 1995; Greshake et al., 2010; Fröhlich et al., 2013).

The mostly granitic and felsic petrology of the crystalline basement of most of the impact structures is quite different compared to that of Mercury. MESSENGER data indicates that the best analogs for the surface of Mercury are basalts or ultramafic komatiites (Nittler et al., 2011; Stockstill-Cahill et al., 2012; Charlier et al., 2013; Maturilli et al., 2014). Unfortunately, naturally shocked basalts are generally rare (Wright et al., 2011).

Still, a spectral study of shocked felsic material is of interest for our purposes. It gives insight into the spectral behavior of highly shocked impactites and their components. The data are also of use for remote sensing of terrestrial impact sites, where impactites with a felsic mineralogy are common (Wright et al., 2011). Studies of other planetary surfaces than Mercury, can also benefit from results obtained in this study. For example, granitic material occurs as fragments and clasts in lunar samples (e.g., Warren et al., 1983; Jolliff et al., 1999; Shervais and McGee, 1999; Seddio et al., 2015). Glotch et al. (2010) observed evolved lithologies and highly silicic material (potential lunar granite) in the mid-infrared using the Diviner Lunar Radio experiment in four lunar regions including the Aristarchus crater, where Mustard et al. (2012) found high-silica impact melt based on Moon Mineral Mapper data.

On the dominantly basaltic Mars granitoid or felsic materials may occur (e.g., Christensen et al., 2005; Ehlmann and Edwards, 2014; Sautter et al., 2014, 2015). Furthermore, there are indications for felsic, granitoid material on Venus (Müller et al., 2008; Gilmore, 2015).

Earlier reflectance and emission studies of impact melt glass and amorphous phases formed in impacts in the mid-infrared were performed, for example, by Thomson and Schultz (2002), Gucsik et al., (2004), Faulques et al., (2005) and Palomba et al., (2006). Spectra of these samples are dominated by a broad feature in the 8.9 to 10.5 μm range, with only few other features in the mid-infrared. Wright et al., (2011) and Basavaiah and Chavan (2013) analyzed bulk material, from the basaltic Lonar Crater, India. Jaret et al., (2013, 2015a) analyzed individual minerals using infrared microscopy. All results show a spectrum with a dominating feature in the ~9.4 - 10.2 µm wavelength range.

Studies on experimentally shocked natural samples (anorthosite, pyroxenite, basalt and feldspar) were made by Johnson et al., (2002, 2003, 2007, and 2012) and Jaret et al., (2015b). In these studies, a degradation of features (loss of intensity, band shifts with increasing shock pressure) were observed due to decreasing structural order. A transmission study of experimentally shocked feldspars by Ostertag (1983) show similar results.

Similar results with strong features between 9 and 10.5 µm were found by Pollack et al., (1973); Crisp et al., (1990); Nash and Salisbury (1991) and Wyatt et al., (2001), who studied terrestrial glasses in basalts or obsidian.

Several mid-infrared reflectance and emission studies used synthetic glasses as analogues for impact melt glass. Byrnes et al., (2007), and Lee et al., (2010) analyzed synthetic quartzofeldspathic glasses, finding very good correlations between the band position of characteristic dominant features and $SiO_2$ contents or Si/O ratios.

Dufresne et al., (2009); Minitti et al., (2002) and Minitti and Hamilton (2010) obtained comparable results for synthetic glass with basaltic to intermediate composition. Basilevsky et al., (2000); Moroz et al., (2010) and Morris et al., (2000) measured glass from laser pulse experiments with Martian soil analogue JSC Mars-1. The resulting spectra are dominated by a strong single feature in the 9.2-10.5 µm wavelength range.

There are also many studies of melt glasses and highly shocked materials in the visible and near-infrared. Raikhlin et al., (1987), Schulz and Mustard (2004) and Iancu et al. (2011) studied terrestrial impact melt rocks, Johnson and Hörz (2003); Adams et al., (1979) and Bruckenthal and Pieters (1984) experimentally shocked feldspars and enstatite. Moroz et al., (2009) studied synthetic Martian impact melt analogs and Bell et al. (1976) and Stockstill-Cahill et al., (2014) analyzed synthetic lunar glass. A study by Keppler (1992) focused on synthetic silicate glasses with albite and diopside composition.

In this study, provide mid-infrared data in the 7-14 µm range of further impact glasses which were so far not covered by mid-infrared studies. Since the surfaces of many planetary bodies are covered by regolith, we have to expect intimate mixtures of grains and particles of different sizes. Grain sizes affect the spectral properties of materials, so this study covers several grain size fractions (0-25, 25-63, 63-125 and 125-250 µm).

A further question is the identification of spectral parameters, which help to identify glasses in remote sensing spectra, and to distinguish them from crystalline material with similar chemical composition.

## 2. Material and Methods

### 2.1 Samples

In order to get representative spectra, we used several grams of melt glass for each of our analyzed samples. Samples (Table 1) were obtained from the collections of the Institut für Planetologie (Münster), Mistastin samples were provided by M.Zanetti (St.Louis). Samples of Lonar crater, Libyan Desert Glass and Irghizite were obtained from meteorite dealers. The Otting and Polsingen samples were made with material used already in Morlok et al., 2016.

### 2.2 Sample Preparation

For the grain size fractions the bulk sample material was first ground in steel and agate mortars into fine powder. The material was cleaned in acetone and dry sieved for at least one hour into four size fractions: 0-25 µm, 25-63 µm, 63-125 µm and 125-250 µm, using an automatic Retsch Tap Sieve. To remove clinging fines, the larger two fractions were again cleaned with acetone.

For additional analyses using optical and Scanning Electron Microscopy (SEM), we produced thick sections of representative blocks of the samples. The sections were polished according to standard procedures for petrological thin sections, guaranteeing a very flat surface.

### 2.3 Optical Microscopy

Overview images of the polished thick sections under normal light were obtained with a KEYENCE Digital Microscope VHX-500F. Light microscopy provides fast information about the general homogeneity, amorphous character, as well as enables first mineral identification in the samples (Fig.1).

### 2.4 SEM/EDX

For imaging and the chemical characterization of the samples, we used a JEOL 6610-LV Scanning electron microscope equipped with an implemented silicon drift Oxford EDX (Energy Dispersive X-Ray Spectroscopy) system. Images were obtained in the Backscattered Electron modus (BSE) to enhance contrast due to chemical variations (Fig.2). Chemical analyses with EDX were quantified with an ASTIMEX™ standard set for major elements. The calibration was confirmed by re-analyzing the standards in each session. Beam current stability was controlled for each analysis using a Faraday cup. For measurements of the chemical composition, we routinely analyzed areas of 100 x 100 $\mu m^2$ using 90 seconds integration times. A broad beam and shorter integration times are helpful to measure volatile elements correctly. Results are presented in Table 2.

## 2.6 Bi-directional reflectance FTIR-Analyses

Each size fraction was gently placed in aluminum sample cups (1 cm diameter, and 1 mm deep). The surface was flattened with a spatula following a similar procedure described by Mustard and Hayes (1997) to create a uniform surface without preferred grain orientations. For the analyses in the mid-infrared from 2.5-19 µm, we used a Bruker Vertex 70 infrared system with a MCT detector at the IRIS (Infrared and Raman for Interplanetary Spectroscopy) laboratory at the Institut für Planetologie (Münster).

Most measurements were conducted under low pressure ($10^{-3}$ bar). In some cases we observed pore collapse during evacuation even in repeated attempts, resulting in surface distortion. The probable reason for pore collapse is rapidly expanding volatiles when low pressures are reached. Attempts to avoid this happening using longer evacuation times or pre-heating to get rid of adsorbed volatiles were also not successful. So we decided to analyze these sample und ambient pressure (see Table 3). This may have resulted in water and atmosphere-related features overlapping with overtone features in the spectral range below 7 µm but is unlikely to affect longer wavelengths.

To ensure a high signal-to-noise ratio, we accumulated 512 scans for each size fraction. For background calibration a diffuse gold standard (INFRAGOLD™) was applied. In order to emulate various observational geometries of an orbiter, we obtained analyses in a variable geometry stage (Bruker A513) for the MERTIS database. The data presented here were obtained in a specular geometry of 30° incidence (i) and 30° emergence angle (e). Results of further analytical geometries are not presented, since the differences are in the intensities of the spectra, while effects on band positions were not observed in this study.

Band positions for both powder and microscope analyses of the characteristic features were obtained using Origin Pro 8. The wavelength of a specific feature was determined by the position of the strongest reflectance, in the case of the CF we used the position of the lowest reflectance.

For the comparison with remote sensing data in the thermal infrared, emission and reflectance data can be compared using Kirchhoff's law: $\varepsilon = 1 - R$ (R=Reflectance, $\varepsilon$ = Emission) (Nicodemus, 1965). However, in this study, a bi-directional, variable mirror set-up was used. This affects the conversion of the reflectance data to emissivity considerably. Kirchhoff's' Law works well for the comparison of directional emissivity and directional hemispherical reflectance (Hapke, 1993; Salisbury et al., 1994), for a direct comparison of directional emissivity with reflectance by using Kirchhoff's law, the reflected light in all directions has to be collected, using a hemispherical reflectance, where a gold-coated hemisphere (Salisbury et al., 1991a; Thomson and Salisbury, 1993; Christensen et al., 2001; King et al., 2004). This makes direct quantitative comparisons between the data obtained in bi-directional mode with surface emission data difficult. For example, anomalous features heave been observed for the CF. (Salisbury et al., 1991b; Salisbury, 1993; Christiansen et al., 2001). However, similarity in band positions and band shapes, as well as the low amount of sample materials needed for the bi-directional analyses (which is important for rare or pure phases) makes this method useful at least for qualitative studies, when these caveats are kept in mind for direct comparisons with remote sensing data.

The environmental conditions on the surface of Mercury also have to be taken into account. Temperatures on the surface can reach up to 700K, at a very low ambient pressure below a pico-bar (e.g. Benkhoff et al., 2010). In contrast, analyses for this study were conducted at low to ambient pressure, and room temperature. Emissivity studies of olivine, pyroxene and feldspar under conditions similar to the surface of Mercury or the Moon (e.g. Donaldson Hanna et al., 2012; Helbert et al., 2009 and 2013) show significant shifts in band positions and also a decrease in spectral contrast for the RB and TF. This also affects direct comparisons between bidirectional reflectance spectra recalculated to emissivity with the remote sensing data.

The spectral range of interest for the MERTIS database is from 7-14 µm. We present powder spectra from 7-19 µm (Fig. 3a-c), as features of interest can appear at longer wavelengths. The signal of the detector used in our study is less effective at wavelengths above 19 µm, resulting in a low signal-to-noise ratio. Spectra are presented in reflectance, from 0-1. For the spectral range from 2.5-7 µm, where the bands for water are located, we present representative spectra in Fig.4.

For characterization purposes, we used reference spectra from the Arizona State University Thermal Emission laboratory (Christensen et al., 2000) and the Johns Hopkins ASTER laboratory (Baldridge et al., 2009).

## 3. Results

### 3.1. Characterizing the Samples using Optical Microscopy and SEM/EDX

Focus of our chemical analyses was to obtain the composition of the pure glass and that of inclusions separately. Mistastin Melt 2 shows low totals below 98 wt% (Table 2a). Similar low totals were observed for Ries glasses in the present and earlier studies (Morlok et al., 2016). These low totals are probably results of high sample porosities and volatile contents. Chemical analyses usually confirm earlier results (for references see Table 1). For better identification, low analytical totals of mineral inclusion were normalized to 100wt% (Table 2b).

The Thailandite and Moldavite tektite samples have all a transparent appearance in optical microscopy (Fig.1). The Thailandite tektite only shows a small inclusion (~0.5 mm) in the otherwise nearly pure glass. The inclusion consists of rutile and/or $SiO_2$ polymorphs (Fig.1 and Table 2b). SEM images also show very homogeneous material (Fig.2). Optical images of Moldavite reveal dendritic crystal aggregates (Fig.1). Optical images of the Muong Nong sample (Fig.1) show abundant layers of small vesicles, which are typical for this type of tektite (e.g. Koeberl, 1992; Son and Koeberl, 2005). Backscattered Electron (BSE) SEM images confirm the abundant small, empty vesicles (Fig.2). The Libyan Desert Glass has empty vesicles in an otherwise very homogeneous, colorless glass, with only few small inclusions (Fig.1) of a crystalline $SiO_2$-polymorph, probably cristobalite, which is typical for this material (Table 2b) (Barnes and Underwood, 1976; Fröhlich et al., 2013). SEM images of the glassy part also show very homogeneous material (Fig.2).

The El'gygytgyn melt glass shows Schlieren-like structures as well as few mineral inclusions in the optical images and in the same SEM pictures, embedded in a homogenous, transparent glass (Fig.1, 2). Inclusions of quartz, magnetite, and ilmenite were observed by Gurov and Koeberl (2004). The Irghizite impact glass has a brown color, and shows abundant (Koeberl and Fredriksson, 1986) empty vesicles in the optical and SEM images, but no larger inclusions (Fig.1, 2).

Popigai melt glass also shows a brown color in the optical images, with abundant vesicles and also some mineral inclusions in the glassy matrix (Fig.1). This was also observed by Whitehead et al., (2002) and Kettrup et al.,(2003). The inclusions were identified as a $SiO_2$-polmorph with SEM/EDX, easily distinguishable from the glass in the SEM image by the blocky appearance (Fig.2, Table 2b). Kettrup et al., (2003) identified Lechatelierite e and coesite in Popigai glass.

The Mien impact melt glass is very heterogeneous, with abundant inclusions in the glassy matrix (Fig.1). The inclusions were characterized as $SiO_2$-polymorphs and pyroxene by SEM/EDX (Table 2b, Fig.2). This confirms the findings of Maerz (1979), who observed abundant quartz and pyroxene mineral fragments. Vesicles also show signs of remnant fillings (Fig.1).

Of the Dellen sample, only powdered material (size >250 µm) was available, so detailed information about the petrographical context is not available. Deutsch et al. (1992) reported a glassy matrix with crystallites e.g., pyroxene. This is consistent with bulk chemical analyses obtained by SEM/EDX in this study (Tab2b).

The samples from the Mistastin lake impact structure have considerable variations. Mistastin melt 1 shows abundant fragments and vesicles, embedded in a brownish, but translucent glass (Fig.1, 2). Fragments identified by EDX were $SiO_2$-polymorphs (Tab 2b). Mistastin melt 2 has a dense, black obsidian-like groundmass with abundant crystallites (also compare Grieve et al., 1975) (Fig.1). The SEM images also show crystallites in a fine-grained matrix (Fig.2). Further phases observed were $SiO_2$-polymorphs, probably quartz (Table 2b) (also see Marion et al., 2010)

The Lonar crater sample is highly vesicular, and the glassy material contains abundant mineral inclusions, ranging from fragments to clusters of small phases (Fig.1, 2). Identified by SEM/EDX were $SiO_2$-polymorphs, pyroxene, and ilmenite (Table 2b). This confirms findings of Kieffer et al., (1976), Osae et al. (2005), and Wright et al., (2011). Impact melt glass from the Lonar crater has average MgO (3.06wt %) and CaO (6.53wt %) (Table 2a) contents slightly outside the range observed for melt rocks and breccia in Osae et al., 2005 (5.32 wt% MgO and 8.81 wt% CaO). Given the heterogeneity and weathered nature of the material, such divergence is expected.

### 3.2. Bi-Directional Reflectance FTIR

All impact melt glasses and rocks show similar spectral features in the 7-14 µm wavelength range (Table 3, Fig.3a-c): A strong RB dominates between 8.9 and 10.3 µm (Features below 7 µm will be discussed separately in 3.4.1 and Fig.4). The Christiansen Feature (CF), a reflectance minimum, varies from 7.3 µm (Desert Glass) to 8.2 µm (Dellen)(Table 3). In many samples, the CF is located at slightly higher wavelengths in the finest size fraction (0-25 µm) when compared to the coarser ones. The TF usually only occurs in the finest size fraction (0-25 µm) (Table 3).

To aid discussion and interpretation, groupings among the spectra and samples, in addition to groupings based on chemistry and their source material (tektites, impact glass), were identified by comparing CF to the $SiO_2$ content (Figure 5). The position of the CF is well correlated with the $SiO_2$ content of the material (Cooper et al., 2002) (Fig.5). In order to avoid artefacts caused by low analytical totals resulting from high porosity in a few samples, we compare the $SiO_2$ content from EDX analyses normalized to 100 wt% totals.

Most of the impact glasses form a group consisting of the felsic/intermediate samples. This group overlaps with the tektites. Glasses from Mistastin and Lonar, while chemically still at the lower end of the intermediate range (>52 wt% $SiO_2$, Le Maitre, 2002), form the (in relative terms) 'mafic' group.

Silicate glasses show an asymmetric stretching of the Si-O bonds in the 8 – 12.5 μm region, overlapping with an asymmetric (Si, Al)-O stretching mode near 9 μm. The specific bands are difficult to distinguish in glass, which results in the characteristic broad band for amorphous silicates. Further Al-O, Si-O-Si and O-Si-O vibrations occur at longer wavelengths over 11 μm (McMillan et al., 1998; King et al., 2004, Dufresne et al., 2009, Speck et al., 2011).

In more detail, the group of tektites (Table 3, Fig.3a) shows similar spectral characteristics, a CF between 7.5 and 7.7 μm, a strong RB from 9.1 to 9.2 μm, and a weak TF between 11.5 and 11.62 μm. The spectra are very smooth, without significant spectral features of crystalline species. The intensities of the strongest RB is also very similar at ~0.1 Reflectance (Fig.3a). Also, there are no significant features at longer wavelengths, except for a weak feature between 12.8 and 13 μm.

The bulk of melt glasses and rocks of the felsic or intermediate groups show a greater spectral variation that also overlaps with those of tektites (Table 3, Fig3b). El'gygytgyn, Irghizite, Popigai, Mien, Libyan Desert Glass, Otting, and Polsingen, show simple spectra with shapes similar to tektites (Fig.3a,b). The spectrum of Libyan Desert Glass is flat at the longer wavelengths as are the spectra of tektites. Most samples show a shoulder between 7.9 – 8.4 μm, especially pronounced in the Libyan Desert Glass (Fig.3a-c). Such shoulders or bands are characteristic for silicate glasses with intermediate composition and related to asymmetric Si-O-Si bridge vibrations (McMillan and Pirou, 1982; Dufresne et al., 2009).

Positions for the characteristic spectral features range from 7.3 to 7.9 μm for the CF, 8.9 to 9.5 μm for the strong RB, and 10.9 to 11.9 μm for the transparency feature (Table 3). Intensities show more variation with the strong RB between 0.05 and 0.1 Reflectance (Fig.3b). Exceptions are the Libyan Desert Glass with the highest intensity in this study, and the Polsingen sample with the lowest intensity (Fig.3b). Libyan Desert Glass is the endmember with these strong RB features occurring at shorter wavelengths, the band positions are also comparable to those of synthetic SiO$_2$ glass (e.g. Faulques et al., 2001). The features in the Ries samples are seen at longer wavelengths (Table 3).

Furthermore, in most samples in this group weak features are observed in the 8.5-8.8 µm range. Red Suevite from Polsingen shows several smaller bands between 8.2 and 8.8 µm and also exhibits stronger bands at longer wavelengths of 17-18.7 µm (Fig.3b), which are probably caused by the crystalline inclusions. These are typical bands of fragments with granite (or similar) composition from unmelted crystalline basement rocks (Baldridge et al., 2009, Morlok et al., 2015). In most cases, at longer wavelengths, there are only weak additional features comparable to the tektites (Fig. 3), probably result of Al-O, Si-O-Si and O-Si-O vibrations (McMillan et al., 1998; King et al., 2004).

The samples from Dellen are an outlier (Fig.5). The CF is at 8-8.2 µm, and the TF at 12.1 µm (Table 3b, Fig3b). The spectra shows no clear main RB, instead it features a 'twin peak', band from 10-10.3 µm which are pyroxene and plagioclase features from fragments (e.g. Baldridge et al., 2009, Hamilton et al, 2000). The strong additional features at longer wavelengths (>17 µm) in Dellen and Polsingen are indicative of weathering phases (e.g. clay; Morlok et al., 2016) (Table 2).

Samples from Mistastin and the Lonar craters form the (relatively) mafic group (Table 3, Fig.3c). The CF falls between 7.9 and 8.1 µm, and the TF is between 11.8 and 12.1 µm. The position of the dominating RB feature ranges from 9.2 to 10 µm, and the intensities of the strongest RB are below 0.1 Reflectance.

However, a second feature observed in Lonar at 10.6 µm (Table 3; Fig.3b) could be a shoulder or feature from unshocked or moderately shocked feldspar (Baldridge et al., 2009; Johnson et al., 2012). Also, comparatively strong features at longer wavelengths (>17 µm; Table 3) hint at weathering phases (Morlok et al., 2016) (Fig.1, 2; Table 2b). In a similar way, the shoulder of the Mistastin Melt 2 spectrum at 8.6-8.7 µm and the features from 17.1-18.6 µm (Table 3) can be explained by the abundant feldspar crystals occurring together with the melt glass, and possibly weathering phases (Fig.1,2) (Grieve et al., 1975; Baldridge et al., 2009; Marion et al., 2010; Morlok et al., 2016).

In general, the strength of the RB feature between 8.9 to 10.3 µm in all spectra shows of this study the dominance of amorphous material in the samples.

### 3.2.1 Water Bands

The detected water bands were normalized to unity in the 2.5-7 µm regions to allow for better comparison of the relative intensity of the water feature (Table 3, Fig.4). Although the position of the water band is very similar in all the samples (it falls between 2.7 and 2.9 µm; e.g. Faulques et al., 2001), the intensities of the feature vary. However, several of the samples have been analyzed under ambient pressure, so they could be influenced by atmospheric water. Tektites, as well as Irghizites and El'gygytgyn show very shallow water bands. Most other samples exhibit broader bands, indicating adsorbed water in addition to OH groups as part of the mineral structure. Libyan Desert Glass has a much sharper feature, indicating a very 'dry' sample.

### 4. Discussion

Results for the strong, dominating RB in impact glasses and tektites in this study (8.9 to 10.3 µm, Table 3) are similar to those obtained in earlier studies, which also exhibit the strong RB band in the 8.9 to 10.5 µm range (Thomson and Schultz, 2002; Gucsik et al., 2004; Faulques et al., 2001; Palomba et al., 2006). In addition, our results are also comparable to terrestrial glasses (9.0-10.5 µm) Pollack et al., (1973); Crisp et al., (1990); Nash and Salisbury (1991) and Wyatt et al., (2001). Series of synthetic glasses based on terrestrial rock compositions (Byrnes et al., 2007; Minitti 2002; DuFresne et al., 2009; Lee et al., 2009; Minitti et al., 2010) show a similar range for this feature from 9.1 to 10.4 µm (for comparisons with the CF see below).

The spectrum of Muong Nong in our study is very similar in band shape and position compared with results obtained by Gucksik et al. (2004): the strongest RB feature is at 9.2 μm, compared to an average of 9.2 μm in this study (Table 3, Fig.3a). An Indochinite spectrum with a strong RB at 9.1 μm (Faulques et al., 2001) shows similarity to both Muong Nong (9.2 μm) and Thailandite (9.1 μm). Another feature near 12.9 μm (Faulques et al., 2001) is found nearby in the results of our study (Table 3). The dominating RB at 9.1 μm for Moldavite and a weaker band at 12.8 μm is close to those in Faulques et al. (2001) for the same tektite (9.0 μm and 12.8 μm) (Table 3). A minor band at 8.1 μm (Faulques et al., 2001) is lacking in our results (Fig. 3a).

The range of results for the main RB in tektites are also comparable to those for synthetic quartzofeldspathic glass with similar composition in Lee et al., (2010), i.e., 9.1-9.2 μm for this study (Table 3) compared to 9.1-9.3 μm. The weak water bands are in good agreement with low water contents (0.002-0.03 wt%) reported in tektites by Beran and Koeberl, 1997 and Dressler and Reimold, 2001) (Fig.4). Other mid-infrared tektite studies are mostly transmission spectra, which cannot be directly compared to our reflectance data. However, the typical shape dominated by a RB feature near 9 μm in tektites is similar to that observed in this study (e.g., Fröhlich et al., 2013, Morlok et al., 2014).

When compared to synthetic glasses with similar composition (Lee et al., 2010), the range in the position for the main RB for the felsic/intermediate group in our study is comparable (Fig.6). Minor features in the 8.5 -8.8 μm and 12.5-12.8 μm ranges observed for most member of this group can be attributed to crystalline quartz inclusions (e.g. Baldridge et al., 2009). Among the felsic and intermediate impact glasses, end member Libyan Desert Glass has very similar features like FTIR analyses by Faulques et al., (2001), which show RB at 7.9, 8.9 μm, and 12.8 μm (Fig.3b). In contrast, the spectra of Gucsik et al., (2004) has only one RB at 9.6 μm, and another weaker feature at 14.4 μm. These differences are indicative of the compositional heterogeneity of the Libyan Desert Glasses.

In the felsic and intermediate group, Irghizite and El'gygytgyn samples have weak water bands like previously seen in the tektites (Beran and Koeberl, 1997) (Fig.4). This indicates very high temperatures during their formation that dehydrated the material, or low abundance of water in the starting material. The stronger water bands of the remaining samples in this group could either point toward higher water contents in the target material, or the influence of weathering. This was also observed for impactites from the Nördlinger Ries crater, where over 2 wt % $H_2O$ was observed in some glasses (Vennemann et al., 2001; Morlok et al., 2016). Similarly, elevated water (or volatile) contents of > 1 wt% were found in melt rock and glass from Dellen (Deutsch et al., 1992), Mien (Schmidt et al., 1997) and Popigai (Kettrup et al., 2003). In addition, the elevated water content of the Libyan Desert Glass compared with tektites was also observed by Faulques et al., (2001).

Results from Dellen samples are an outlier. Based on its $SiO_2$ content, the RB and CF should be at shorter wavelengths (Fig.5). However, the positions are not surprising given the number of crystalline inclusions in the glass. For example, the observed high abundance of pyroxenes (Table 2) could have moved the band position of the feature to longer wavelengths (Salisbury, 1991a; Deutsch et al., 1992).

In the more 'mafic' group (compared to the previously discussed felsic/intermediate group), the spectrum of the Lonar samples exhibit the strongest RB at 9.2 µm (Table 3), similar to findings of Wright et al., (2011), which have the RB at 9.4-9.5 µm for samples showing transformation to maskelynite (class 2) to complete melting (class 5) (Kieffer et all, 1976; Wright et al., 2011) (Fig.3c). Jaret et al. (2015a) identified a similar CF position at ~8.0 µm (8.0-8.1 µm in this study, Table 3). On the other hand, in-situ analyses of completely amorphous material from Lonar have the main feature at 10-10.4 µm (depending on grain orientation), while a feature at 9.3 µm occurs in spots which retained some crystallinity (Jaret et al., 2013, 2015a). Differences between the spectra can be explained with the highly heterogeneous character of the materials resulting in high contents of fragments (Fig.1, 2) (Kieffer, 1976; Osae et al., 2005; Wright et al., 2011). The normalized water feature of the Lonar material is comparatively weak, the intensity is located between that of most samples and the tektites (Fig.4).

The two Mistastin melts show slightly different spectra (Fig.3c), likely due to the degree of crystallinity in the different samples. Mistastin Melt 2 has additional features (8.6-8.7 µm, 17.1-18.6 µm (Table 3) explained by the abundant crystals within the melt glass (Fig.1,2) (Grieve et al., 1975; Marion et al., 2010). Mistastin Melt 1, on the other hand, has a spectrum typical for entirely amorphous materials (e.g. Lee et al., 2010). The position of the strong RB (9.6-9.7 µm, Table 3) is similar to that of synthetic glasses (~9.8 µm, Lee et al., 2010). The water bands are comparable to those for the other studied melt glasses (Fig. 4).

### 4.1. Identification of Glass

Glasses are characterized in the context of this study by a broad main RB in the 8.9 to 10.3 µm range (Table 3), in contrast to the usually more feature-rich crystalline felsic and intermediate rocks (Salisbury, 1988). A comparison of this dominant RB with $SiO_2$ content shows similar characteristics like the comparison of the CF and the $SiO_2$ content (Fig.6). With increasing $SiO_2$ content, the band position of the RB moves to shorter wavelengths (also compare Cooper et al., 2002; Lee et al., 2010). However, in some cases remote sensing data only provides spectra with weak RB features e.g. in many ground based telescope observation in the mid-infrared of Mercury (e.g. Tyler et al., 1988, Emery et al., 1998, Cooper et al., 2001, Sprague et al., 2002, 2007). So, additional spectral features to identify glassy material are of interest.

If the CF as reflectance minimum (or emission maximum) is obtained in such observations, it is comparatively easy to identify in remote-sensing data. In a direct comparison between CF and $SiO_2$-content (Fig.5), the results of our study plot on a slope very similar to that observed for a series of powdered crystalline plutonic rocks in earlier works (Cooper et al., 2002), with only the exception of the Dellen and the Libyan Desert Glass data (Fig.5). Thus, on the basis of this type of diagram, it would be difficult to distinguish melt glass from crystalline material. The independence of the CF from degree of crystallinity was already observed e.g. for feldspar by Nash and Salisbury, 1990.

An alternative is the comparison between the CF and the SCFM ($SiO_2$, CaO, FeO, MgO) index. The abundance of divalent cations Ca, Fe, and Mg affects the depolymerization of the silicate tetrahedra. The SCFM index is calculated from the oxides of Si, Ca, Fe, and Mg: $SiO_2/(SiO_2+CaO+FeO+MgO)$ (Walter and Salisbury, 1989; Cooper et al., 2002). The results (Fig.7) show essentially a similar grouping as in Fig.5. However, impact melt glasses differ slightly from the trend line for powdered crystalline rocks with comparable chemical composition using the SCFM index (Fig.7), with most results plotting below the trend line. The SCFM is sensitive to the degree of polymerization of the rocks (Cooper et al., 2002), while glasses have a low degree of ordering with only short range structures present. This could explain the divergence, and so SCFM values plotting below the line for crystalline rocks are possibly indicative of melt glasses.

However, in a remote sensing situation the divergence would require additional chemical data for the observed material in order to differentiate between glass and crystalline material. Chemical data for surface regions on Mercury was calculated form data obtained by the X-Ray Spectrometer (XRS) and the Gamma-Ray Spectrometer (GRS) on the MESSENGER probe that orbited Mercury (Charlier et al., 2013; Stockstill-Cahill et al., 2013; vander Kaaden et al., 2015; Peplowski et al., 2015). The BepiColombo mission will include MGNS (Mercury Gamma-ray and Neutron Spectrometer) and MIXS (Mercury Imaging X-ray Spectrometer) which will provide chemical data in direct alignment with the mid-infrared data from MERTIS (Benkhoff et al., 2010).

A comparison with micro-FTIR studies of synthetic glass of felsic, intermediate, and basaltic composition (Fig.6) (Lee et al., 2009) shows a difference between the natural impact melt glass and the synthetic material. Since the general chemistry of the natural and synthetic glasses is comparable, effects of different analytical techniques or sample type cannot be ruled out. Cooper et al. (2002) observed differences between analyses of powdered and solid rocks. On the other hand, Klima and Pieters (2006) observed no differences regarding band positions between these types of material.

Additional spectral information in the near-infrared could also help to distinguish between crystalline and glassy material (e.g. Gaffey et al., 1993). In the case of remote sensing of Mercury, such data is also available from the MESSENGER mission, and will also be provided by BepiColombo with the Visible Infrared Hyperspectral Imager Channel (VIHI) on the Spectrometer and Imagers for BepiColombo Integrated Observatory System (SIMBIO-SYS) instrument suite (Benkhoff et al., 2010).

**4.2. Application to MERTIS and Remote Sensing of Mercury**

Mid-infrared data from Mercury is rare due to difficulties for ground-based observations (e.g. Cooper et al., 2001). Spectral studies by ground- and airplane based telescopes in the mid-infrared were made by Sprague et al. 1994, 2000, 2002, 2007; Emery et al. 1998; Sprague and Roush 1998; Cooper et al. 2001; Donaldson-Hanna et al. 2007). Due to observational limitations, they all cover large surface areas from at least $10^4$ - $10^6$ km$^2$. The dominating mineral phases identified in these studies are mainly plagioclase with minor pyroxene (e.g. Sprague et al., 2007).

Most spectra from Mercury show only weak features and low spectral contrast and probably have a low signal to noise ratio. Potential CF are visible from 7.7 – 8.7 μm. The CF observed in this study overlap with this range at shorter wavelengths, with the glasses from the 'mafic' group, Dellen and some from the intermediate group falling into the range observed for the CF on Mercury (Tab.3). The TF for the observations of Mercury is between 12 and 12.7 μm, only partially overlapping with the TF in our study (10.9 – 12.1 μm). Only Dellen, Mistastin and Lonar samples have CF in this region (Tab.3) (Sprague et al. 1994, 2000, 2002, 2007; Emery et al. 1998; Sprague and Roush 1998; Cooper et al. 2001; Donaldson-Hanna et al. 2007). RB are difficult to identify in the surface spectra. Candidates are found mostly at 9.2-9.5 μm (Emery et al., 1998;

Sprague et al., 1994; Sprague et al., 1998; Sprague et al., 2002; Donaldson-Hanna 2007) would overlap with most of the strong RB in the intermediate and mafic group (Tab.3).

Since these areas observed so far are from vast surface regions, the spectral features of distinct regions with characteristic compositions are integrated into one spectrum. This makes it difficult to discuss the results in a wider context, such as smaller surface features. Here the high spatial resolution of MERTIS of ~500 m (Benkhoff et al., 2010) will allow to resolve smaller structures such as the hollows (e.g. Thomas et al. 2014).

However, regolith is a mixture of many phases, so glass would be only one of several mineral/glass components and size fractions, which have to be taken into account in spectral deconvolution modelling. Also, the conversion problems (See 2.6 Bi-directional reflectance FTIR-Analyses) of bi-directional reflectance for comparison with emissivity data will have to be taken into account for direct comparisons of laboratory and remote sensing data.

## 5. Summary and Conclusions

We have characterized and analyzed mid-infrared spectra of four size fractions (0-25, 25-63, 93-125 and 125-250 μm) from 14 impact melt glass samples, covering the compositional range from highly felsic (Libyan Desert Glass) to mafic/basaltic (Lonar crater melt).

Most samples show mid-infrared spectra typical of highly amorphous material, dominated by a strong Reststrahlen Band between 8.9 and 10.3 μm (Table 3). Even substantial amounts of mineral fragments hardly affect this general band shape. However, crystallization from the melt, such as in one of our two Mistastin samples (Mist Melt 2), can result in significant differences in the spectra between compositionally similar samples.

An additional sign of a highly amorphous state is the lack of features at wavelengths longer than ~15 μm (Fig.3a-c). Tektites have much weaker water features than most of the other impact melt glasses, with the exception of samples from Irghizite and El'gygytgyn.

For the application in remote sensing, spectral features have to be correlated with compositional characteristics of the materials. A convenient method to correlate compositional and spectral features is to use the prominent and in spectra with low signal/noise ratio easily recognizable Christiansen Feature and compare it to compositional parameters like the SCFM index (Cooper et al., 2002). The comparison shows some differences, which could help distinguish glass from crystalline material in remote sensing, using chemical compositional data from further instruments onboard of BepiColombo.

A comparison between the laboratory spectra and mid-infrared ground-observations of Mercury shows similarities in band positions between various features. However, there is no direct 'fit', the occurrence of glasses similar to those analyzed in our study on the surface of Mercury has to be confirmed by more detailed studies including a wider range of phases.

**Acknowledgements**

This work is supported by the DLR funding 50 QW 1302 in the framework of the BepiColombo mission. Also many thanks to Alexander Deutsch (Münster) and Mike Zanetti (St.Louis) for providing the samples.

**Figure Captions**

Figure 1. Optical images of polished sections from the studied samples. Tektites Thailandite, Muong Nong and Moldavite can be recognized by their transparent appearance and very low content of inclusions. Most other melt glasses show higher abundances of crystalline material e.g. Popigai, Mien), while Libyan Desert glass has an appearance similar to the tektites. Samples like Lonar or Mistastin Melt 2 have very high contents in inclusions.

Figure 2. SEM/BSE images of polished sections from the samples analyzed in this study. The findings of the optical microscopy are mainly confirmed, high contents of vesicles are visible in the Muong Nong, Irghizite and Lonar samples.

Figure 3. Mid-infrared bi-directional reflectance spectra of impact melt and glass samples. Band positions in µm. In reflectance (0-1). Blue: 0-25 µm, Pink: 25-63 µm, Red: 63-125 µm, Brown: 125-250 µm (in µm). (a) Tektites, (b) Felsic/intermediate Samples, (c) Mafic/basaltic samples.

Figure 4. Spectral range from 2.5 to 7 µm for each sample (always size fraction 125-250µm, normalized on unity), which shows the water features at ~3 µm. Spectra are offset along the y-axis for clarity.

Figure 5. Comparison of $SiO_2$ concentration (in wt%) in the samples with the position of the Christiansen Feature (in µm). Dotted line: Crystalline felsic and intermediate rocks from Cooper et al. (2002). Most samples from this study form a group of felsic/intermediate material (including the tektites), with a small group of near-mafic samples (Mistastin, Lonar) forming the 'mafic' group.

Figure 6. Comparison of $SiO_2$ concentration (in wt%) in the samples with the position of the strongest Reststrahlen Band (in µm). The dotted line is the trend line for the synthetic glass with intermediate to felsic composition (Lee et al., 2010). Most results for the glasses differ from the results for synthetic glass (DuFresne et al., 2009; Lee et al., 2010).

Figure 7. Comparison of the SCFM-index ($SiO_2/(SiO_2+CaO+FeO+MgO)$) (Walter and Salisbury, 1989, Cooper et al., 2002) with the position of the Christiansen Feature (in μm). Dotted line: Crystalline felsic and intermediate rocks from Cooper et al., (2002). The results for the impact glass mostly plot below the trend line for crystalline material.

| Source | Sample | Age (Ma) | Size crater (km) | Source rock | Lit. |
|---|---|---|---|---|---|
| Australasian Strewnfield | Muong Nong | 0.8 | 90-116? | Marine/River sediments? | [1,2,3,4] |
| Australasian Strewnfield | Thailandite/ | 0.8 | 90-116? | Marine/River sediments? | [1,2,3,4] |
| Nördlinger Ries Crater | Moldavite | 14.3 | 26 | Sediments, sands, clays | [5,6,7,8,9] |
|  | Otting Polsingen | 14.3 | 26 | Granite, Gneiss, Amphibolite | [10] |
| El'gygytgyn Impact Structure | El'gygytgyn | 3.58 | 18 | Rhyodacitic/Rhyolitic Ignimbrite Basalt, Andesite | [11,12,13] |
| Zhamanshin Impact Structure | Irghizite | 0.9 | 5.5-6.3 | Quartz-sericite schist, phyllite , clays and locally sandstones, Ultrabasic Intrusion | [6,14,15, 16,17] |
| Popigai Impact structure | Popigai | 35.7 | 100 | Gneiss | [18,19,20] |
| Mien Impact Structure | Mien | 118.7 | 9 | Granite Gneiss Amphibole | [21,22] |
| ? | Libyan Desert Glass | 29 | ? | Sands/Sandstone | [5,23] |
|  |  |  |  |  |  |
| Dellen Impact Structure | Dellen | 89 | 20 | Granodiorite Gneiss | [21,24] |
|  |  |  |  |  |  |
| Mistastin Lake Impact Structure | Mistastin | 36 | 28 | Granodiorite, Anorthosite, Mangerite | [25,26,27,28] |
| Lonar Lake Crater | Lonar | 0.57 | 1.8 | Basalt | [29,30,31] |

Table 1. Overview of the samples and their sources used in this study. Age in million years (Ma), Crater size in km, Basement rock: petrology of probably source material for melts. Lit.: literature sources for data: [1] French 1998, [2] Chaussidon and Koeberl 1995, [3] Izett and Obradovich 1992, [4] Lee and Wei 2000, [5] Magna et al., 2011, [6] Koeberl and Fredriksson., 1986, [7] Laurenzi et al., 2003, [8] Buchner et al., 2003, [9] von Engelhardt et al., 2005, [10] Stöffler et al., 2013, [11] Layer et al., 2000, [12] Raschke et al., 2014, [13] Gurov and Koeberl 2004, [14] Koeberl , 1986,[15] Taylor et al., 1979, [16] Ackerman et al., 2015, [17] Deino et al., 1990, [18] Bottomley et al., 1997, [19] Withehead et al., 2002, [20] Kettrup et al., 2003, [21] Schmidt et al., 1997 [22] Bottomley et al., 1978, [23] Fröhlich et al., 2013, [24] Deutsch et al.,1992, [25] Marion et al., 2010, [26] Grieve 1975, [27] Marchand et al., 1977, [28] Pickersgill et al., 2015, [29] Osae et al., 2005, [30] Wright et al., 2011, [31] Jourdan et al., 2001.

Table 2a. Chemical Composition of the impact melt rocks and glasses. SEM/EDX data, in wt%. Results for Otting and Polsingen are from Morlok et al., 2016. s.d.=Standard Deviation (1Ω). Mist.=Mistastin. Elgy.= El'gygytgyn.

| | Muong Nong | s.d. | Thai-landite | s.d. | Molda-vite | s.d. | Elgy | Irghizite | s.d. | Popigai | s.d. | Mien | Desert Glass | s.d. | Otting | s.d. | Polsingen | s.d. | Dellen | s.d. |
|---|---|---|---|---|---|---|---|---|---|---|---|---|---|---|---|---|---|---|---|---|
| $Na_2O$ | 1.39 | ±0.01 | 1.24 | ±0.01 | 0.43 | ±0.02 | 2.92 | 1.03 | ±0.02 | 2.45 | ±0.03 | 3.18 | 0.03 | ±0.02 | 2.79 | ±0.27 | 3.23 | ±0.31 | 2.83 | ±0.57 |
| MgO | 1.96 | ±0.12 | 1.96 | ±0.02 | 1.53 | ±0.04 | 0.82 | 2.92 | ±0.04 | 4.00 | ±0.02 | 0.38 | 0.01 | ±0.01 | 2.82 | ±0.26 | 0.33 | ±0.14 | 0.20 | ±0.01 |
| $Al_2O_3$ | 14.80 | ±0.20 | 13.17 | ±0.09 | 9.82 | ±0.08 | 15.25 | 9.51 | ±0.08 | 16.90 | ±0.10 | 14.83 | 1.15 | ±0.63 | 14.99 | ±0.42 | 17.91 | ±0.75 | 12.09 | ±0.25 |
| $SiO_2$ | 70.65 | ±0.22 | 72.70 | ±1.00 | 79.81 | ±0.59 | 70.79 | 73.66 | ±0.37 | 60.78 | ±0.22 | 69.82 | 99.07 | ±1.19 | 59.84 | ±1.10 | 56.37 | ±0.84 | 74.56 | ±0.37 |
| $P_2O_5$ | n.d. | | n.d. | | n.d. | | 0.08 | 0.06 | ±0.04 | 0.12 | ±0.03 | 0.16 | n.d. | | 0.39 | ±0.04 | 0.46 | ±0.08 | 0.24 | ±0.04 |
| $SO_3$ | 0.09 | ±0.06 | 0.06 | ±0.04 | 0.08 | ±0.07 | 0.10 | 0.05 | ±0.04 | 0.35 | ±0.08 | 0.09 | 0.13 | ±0.02 | 0.06 | ±0.02 | 0.06 | ±0.04 | 0.03 | ±0.02 |
| $K_2O$ | 2.84 | ±0.02 | 2.44 | ±0.06 | 3.43 | ±0.03 | 4.17 | 1.91 | ±0.01 | 2.94 | ±0.02 | 5.29 | n.d. | | 3.98 | ±0.25 | 5.85 | ±0.59 | 4.85 | ±0.41 |
| CaO | 1.33 | ±0.15 | 1.89 | ±0.05 | 1.97 | ±0.06 | 2.64 | 2.38 | ±0.01 | 2.97 | ±0.03 | 1.83 | n.d. | | 3.65 | ±0.25 | 3.37 | ±0.42 | 0.94 | ±0.17 |
| $TiO_2$ | 0.86 | ±0.03 | 0.76 | ±0.04 | 0.37 | ±0.02 | 0.35 | 0.76 | ±0.04 | 0.87 | ±0.03 | 0.40 | 0.17 | ±0.06 | 0.93 | ±0.05 | 1.03 | ±0.05 | 0.49 | ±0.02 |
| $Cr_2O_3$ | 0.04 | ±0.02 | 0.02 | ±0.02 | 0.02 | ±0.02 | 0.05 | 0.06 | ±0.03 | 0.05 | ±0.03 | n.d. | 0.01 | ±0.01 | 0.04 | ±0.01 | 0.02 | ±0.01 | n.d. | |
| MnO | 0.11 | ±0.01 | 0.11 | ±0.05 | 0.05 | ±0.01 | 0.08 | 0.09 | ±0.03 | 0.07 | ±0.01 | 0.03 | 0.02 | ±0.02 | 0.09 | ±0.03 | 0.01 | ±0.01 | 0.04 | ±0.01 |
| FeO | 5.45 | ±0.02 | 4.79 | ±0.05 | 1.73 | ±0.03 | 3.03 | 5.97 | ±0.08 | 8.00 | ±0.02 | 1.99 | 0.10 | ±0.04 | 4.81 | ±0.29 | 2.14 | ±0.64 | 2.55 | ±0.11 |
| NiO | 0.03 | ±0.04 | 0.01 | ±0.03 | 0.01 | ±0.02 | 0.00 | 0.18 | ±0.02 | 0.00 | ±0.00 | 0.00 | 0.02 | ±0.02 | 0.01 | ±0.01 | 0.00 | ±0.01 | 0.04 | ±0.03 |
| SUM | 99.56 | | 99.14 | | 99.24 | | 100.28 | 98.58 | | 99.50 | | 98.00 | 100.70 | | 94.41 | | 90.76 | | 98.83 | |

Table 2 cont.

| | Mist. Melt 1 | s.d. | Mist. Melt 2 | Lonar | s.d. |
|---|---|---|---|---|---|
| $Na_2O$ | 4.13 | ±0.06 | 3.99 | 1.28 | ±0.13 |
| MgO | 1.22 | ±0.02 | 1.03 | 3.08 | ±0.18 |
| $Al_2O_3$ | 20.35 | ±0.08 | 20.56 | 14.67 | ±1.40 |
| $SiO_2$ | 56.61 | ±0.04 | 53.54 | 57.59 | ±1.38 |
| $P_2O_5$ | 0.46 | ±0.01 | 0.41 | 0.65 | ±0.05 |
| $SO_3$ | 0.04 | ±0.01 | 0.11 | 0.06 | ±0.05 |
| $K_2O$ | 1.93 | ±0.05 | 1.51 | 1.62 | ±0.00 |
| CaO | 7.18 | ±0.06 | 6.82 | 6.53 | ±0.42 |
| $TiO_2$ | 1.10 | ±0.00 | 1.09 | 2.02 | ±0.23 |
| $Cr_2O_3$ | 0.07 | ±0.03 | 0.05 | 0.02 | ±0.02 |
| MnO | 0.10 | ±0.01 | 0.05 | 0.16 | ±0.06 |
| FeO | 5.99 | ±0.14 | 5.55 | 11.67 | ±0.88 |
| NiO | 0.02 | ±0.02 | 0.02 | n.d. | |
| SUM | 99.16 | | 94.73 | 99.31 | |

|  | Thai Rutile | SiO$_2$ | Elgy. | Popigai SiO$_2$ | Mien Pyx | SiO$_2$ | Libyan Desert SiO$_2$ | Dellen | Mist. Melt 1 SiO$_2$ | Melt. 2 SiO$_2$ | Lonar Pyx | SiO$_2$ | Ilmenite |
|---|---|---|---|---|---|---|---|---|---|---|---|---|---|
| Na$_2$O | 0.09 | 0.04 | 2.88 | 0.08 | 0.08 | 0.05 | 0.07 | 0.09 | 0.05 | 0.06 | 0.15 | 0.13 | 0.09 |
| MgO | 0.01 | 0.03 | 0.93 | 0.03 | 25.80 | 0.07 | n.d | 17.56 | 0.02 | 0.03 | 15.91 | 0.01 | 1.25 |
| Al$_2$O$_3$ | 0.47 | n.d | 15.15 | 0.26 | 1.00 | 0.64 | 1.69 | 1.49 | n.d. | n.d | 0.94 | 1.69 | 0.16 |
| SiO$_2$ | 0.87 | 99.19 | 70.47 | 99.71 | 54.21 | 98.80 | 98.26 | 50.93 | 98.95 | 99.26 | 50.43 | 95.92 | 0.19 |
| P$_2$O$_5$ | n.d | 0.05 | 0.07 | 0.06 | n.d | 0.03 | 0.12 | 0.03 | 0.11 | 0.06 | 0.07 | 0.03 | 0.07 |
| SO$_3$ | n.d | 0.04 | n.d | 0.11 | 0.07 | 0.06 | 0.09 | 0.04 | 0.06 | 0.11 | 0.06 | 0.09 | 0.02 |
| K$_2$O | 0.03 | n.d | 4.07 | n.d | 0.01 | n.d | n.d | 0.01 | n.d | n.d | n.d | 0.71 | n.d |
| CaO | 0.17 | n.d | 2.74 | 0.03 | 1.09 | n.d | n.d | 0.93 | n.d | n.d | 4.69 | 0.38 | 0.14 |
| TiO$_2$ | 97.97 | n.d | 0.39 | 0.04 | 0.16 | 0.03 | 0.21 | 0.27 | 0.03 | 0.01 | 0.58 | n.d | 51.77 |
| Cr$_2$O$_3$ | n.d | 0.03 | 0.05 | 0.01 | 0.21 | 0.00 | 0.02 | 0.05 | 0.02 | n.d | n.d | 0.03 | 0.23 |
| MnO | n.d | n.d. | 0.10 | n.d | 0.59 | 0.06 | n.d | 0.54 | 0.01 | 0.06 | 0.52 | n.d | 0.48 |
| FeO | 1.51 | 0.57 | 3.11 | 0.13 | 17.56 | 0.25 | 0.15 | 29.79 | 0.05 | 0.03 | 27.17 | 0.05 | 45.53 |
| NiO | n.d | 0.06 | 0.06 | 0.03 | 0.19 | n.d | n.d | 0.11 | 0.01 | n.d | n.d | n.d | 0.09 |
| SUM | 101.12 | 100.00 | 100.00 | 100.49 | 100.97 | 100.00 | 100.61 | 101.84 | 99.31 | 99.62 | 100.52 | 99.04 | 100.00 |

Table 2b. Chemical Composition for some inclusions in the impact melts and glass. SEM/EDX data, in wt%. Samples with low totals below 98%wt% were normalized to 100wt% for better mineral identification. Mist.=Mistastin, Elgy= El'gygytgyn, Libyan Desert=Libyan Desert Glass, Thai=Thailandite, Pyx=Pyroxene, Qtz=Quartz, Ilm=Ilmenite. Mist.=Mistastin. Elgy.= El'gygytgyn.

|  | CF |  |  |  | TF |  |
|---|---|---|---|---|---|---|
| **Thailandite** | | | | | | |
| **0-25** | 2.8 | 5.42 | 7.65 | 9.13 | 11.6 | |
| **25-63** | 2.79 | 5.42 | 7.63 | 9.12 | | 12.83 |
| **63-125** | 2.79 | 5.42 | 7.56 | 9.13 | | 12.78 |
| **125-250** | 2.79 | 5.45 | 7.63 | 9.13 | | 12.79 |
| | | | | | | |
| **Muong Nong** | | | | | | |
| **0-25** | 2.8 | 5.4 | 7.68 | 9.16 | 11.62 | |
| **25-63** | 2.8 | 5.4 | 7.67 | 9.15 | | 12.97 |
| **63-125** | 2.8 | 5.4 | 7.67 | 9.15 | | 12.82 |
| **125-250** | 2.8 | 5.5 | 7.64 | 9.15 | | 12.82 |
| | | | | | | |
| **Moldavite** | | | | | | |
| **0-25** | 2.79 | 5.42 | 7.55 | 9.07 | 11.54 | |
| **25-63** | 2.79 | 5.41 | 7.54 | 9.07 | | 12.85 |
| **63-125** | 2.79 | 5.41 | 7.53 | 9.07 | | 12.79 |
| **125-250** | 2.79 | 5.42 | 7.53 | 9.07 | | 12.79 |
| | | | | | | |
| **El'gygytgyn** | | | | | | |
| **0-25** | 2.8 | 5.5 | 7.71 | 9.24 | 11.62 | |
| **25-63** | 2.8 | 5.52 | 7.66 | 9.24 | | 12.89 |
| **63-125** | 2.8 | 5.53 | 7.64 | 9.24 | | 12.88 |
| **125-250** | 2.8 | 5.57 | 7.64 | 9.25 | | 12.99 |
| | | | | | | |
| **Irghizite** | | | | | | |
| **0-25** | 2.79 | 5.4 | 7.61 | 9.09 | 11.6 | |
| **25-63** | 2.79 | 5.4 | 7.6 | 9.08 | | 12.77 |
| **63-125** | 2.79 | 5.4 | 7.61 | 9.08 | | 12.77 |
| **125-250** | 2.79 | 5.43 | 7.61 | 9.09 | | 12.75 |

Table 3. Band positions of features in powdered impact melts and glasses (in µm). CF=Christiansen Feature, TF=Transparence Feature, bel.: analyzed under ambient pressure.

|  | CF |  |  |  |  |  |  |  | TF |  |  |  |  |
|---|---|---|---|---|---|---|---|---|---|---|---|---|---|
| **Popigai** | | | | | | | | | | | | | |
| **0-25** | 2.8 | 5.4 | 6.15 | 7.72 | | 8.56 | | 9.21 | | 11.45 | 13.36 | 14.81 | |
| **25-63** | 2.78 | 5.4 | 6.15 | 7.65 | | | | 9.24 | | | 12.54 | 12.84 | |
| **63-125** | 2.78 | 5.4 | 6.15 | 7.6 | | | | 9.21 | | | 12.54 | 12.8 | |
| **125-250** | 2.78 | | 6.15 | 7.62 | | | | 9.32 | | | | 12.82 | |
| **Mien (bel.)** | | | | | | | | | | | | | |
| **0-25** | 2.75 | | 6.1 | 7.78 | | | | 9.17 | | 11.75 | | 17-17.5 | 18-19 |
| **25-63** | 2.75 | | 6.1 | 7.73 | | 8.58 | | 9.17 | | 11.71 | | 17-17.5 | 18-19 |
| **63-125** | 2.75 | | 6.1 | 7.61 | | | | 9.15 | | | | 17-17.5 | 18-19 |
| **125-250** | 2.75 | | 6.1 | 7.59 | | | | 9.16 | | | | 17-17.5 | 18-19 |
| **Libyan Desert Glass** | | | | | | | | | | | | | |
| **0-25** | 2.72 | 5.34 | 6.17 | 7.28 | 7.92 | | | 8.9 | | 11 | | | |
| **25-63** | 2.72 | 5.34 | 6.15 | 7.29 | 8 | | | 8.89 | | 10.92 | | | |
| **63-125** | 2.72 | 5.34 | 6.16 | 7.28 | 8.03 | | | 8.89 | | | | | |
| **125-250** | 2.72 | 5.44 | | 7.27 | 8.09 | | | 8.9 | | | | | |
| **Otting Glasbombe (bel.)** | | | | | | | | | | | | | |
| **0-25** | 2.75 | | 6.1 | 7.82 | | 8.57 | | 9.33 | | 11.76 | | | |
| **25-63** | 2.75 | | 6.1 | 7.8 | | 8.58 | | 9.34 | | | | | |
| **63-125** | 2.75 | | 6.1 | 7.79 | | | | 9.39 | | | | | |
| **125-250** | 2.75 | | 6.1 | 7.77 | | | | 9.38 | | | | | |
| **Polsingen (bel.)** | | | | | | | | | | | | | |
| **0-25** | 2.76 | 2.9 | 6.1 | 7.88 | 8.22 | 8.52 | 8.72 | 9.48 | | 11.81 | | | |
| **25-63** | 2.76 | 2.9 | 6.1 | 7.84 | 8.27 | 8.56 | 8.78 | 9.47 | | 11.89 | | 17.0-17.5 | 18.7 |
| **63-125** | 2.76 | 2.9 | 6.1 | 7.76 | 8.32 | 8.57 | 8.84 | 9.48 | | | | 17.0-17.5 | 18.7 |
| **125-250** | 2.76 | 2.9 | 6.1 | 7.61 | | | 8.84 | 9.49 | | | | 17.0-17.5 | 18.7 |
| **Dellen3/4 (bel.)** | | | | | | | | | | | | | |
| **0-25** | 2.75 | | 6.1 | 8.19 | | 8.75 | | 9.66 | 10.01 | 10.27 | 12.09 | | |
| **25-63** | 2.75 | | 6.1 | 8.12 | | 8.76 | | 9.69 | 10 | | | 15.6 | |
| **63-125** | 2.75 | | 6.1 | 8.06 | | 8.77 | | | 9.95 | | | 15.7 | |
| **125-250** | 2.75 | | 6.1 | 8.01 | | | | | 9.97 | 10.31 | | 15.68 | |

Table 3 cont.

|  | CF |  |  |  |  |  | TF |  |  |
|---|---|---|---|---|---|---|---|---|---|
| **Mistastin Melt 1** | | | | | | | | | |
| **0-25** | 2.9 | | 7.99 | | 9.66 | | 11.8 | | |
| **25-63** | 2.8 | | 7.96 | | 9.64 | | | | |
| **63-125** | 2.8 | | 7.9 | | 9.66 | | | | |
| **125-250 µm** | 2,8 | | 7,92 | | 9.65 | | | 14.31 | |
| **Mistastin Melt 2 (bel.)** | | | | | | | | | |
| **0-25** | 2.76 | 6.1 | 8.07 | 8.56 | 9.96 | | 12.13 | 17.07 | 18.6 |
| **25-63** | 2.76 | 6.1 | 8.01 | 8.74 | 9.88 | | 12.05 | 17.29 | 18.43 |
| **63-125** | 2.76 | 6.1 | 8.02 | | 9.89 | | | 17.28 | 18.47 |
| **125-250** | 2.76 | 6.1 | 7.93 | | 9.86 | 9.68 | | 17.25 | 18.48 |
| **Lonar Melt (bel.)** | | | | | | | | | |
| **0-25** | 2.74 | 6.1 | 8.14 | | 9.18 | 10.68 | 12.09 | | |
| **25-63** | 2.74 | 6.1 | 8.04 | | 9.19 | 10.62 | | 17.76 | |
| **63-125** | 2.74 | 6.1 | 8.03 | | 9.21 | 10.61 | | 17.76 | |
| **125-250** | 2.74 | 6.1 | 8.01 | | 9.22 | 10.59 | | 17.74 | |

Table 3 cont.

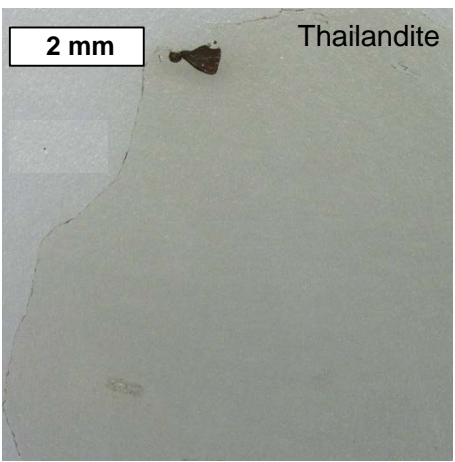
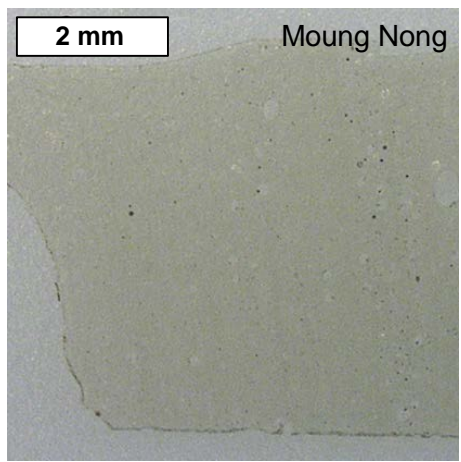
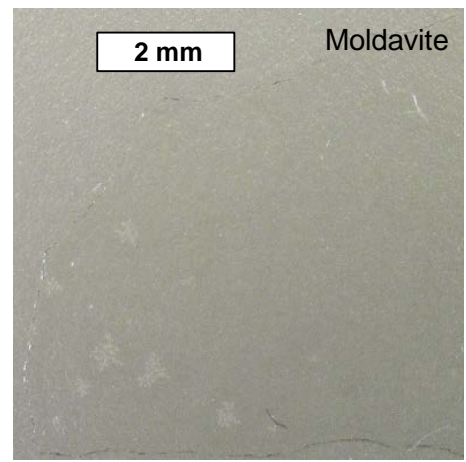
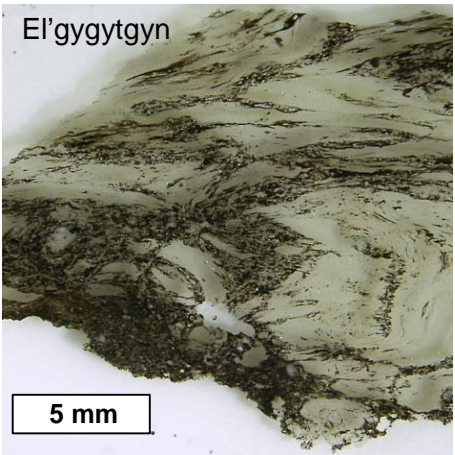
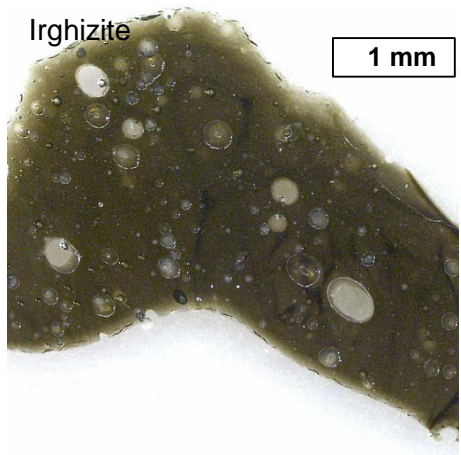
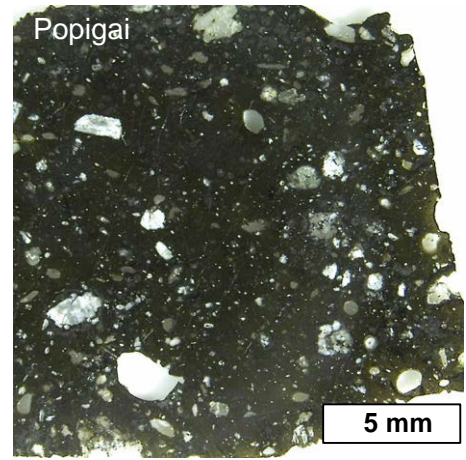
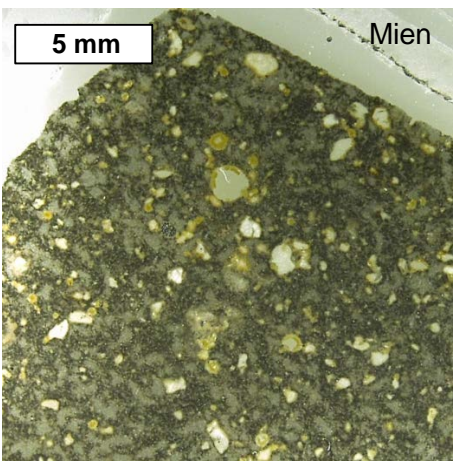
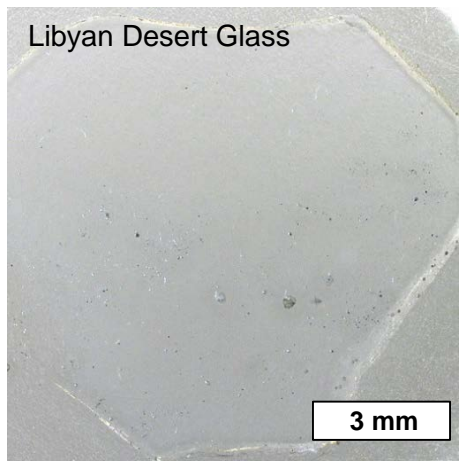
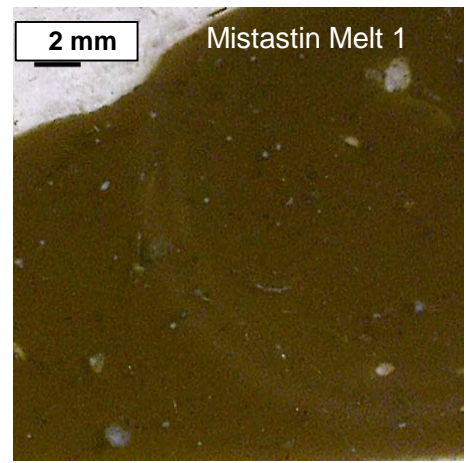
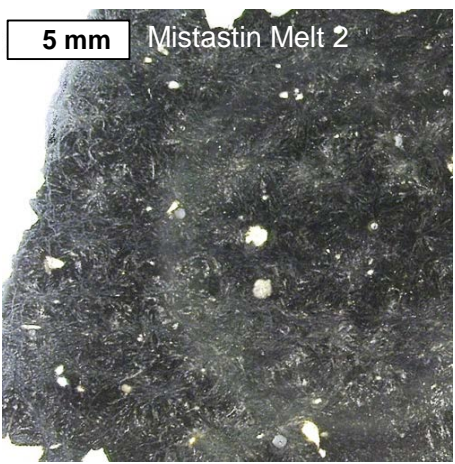
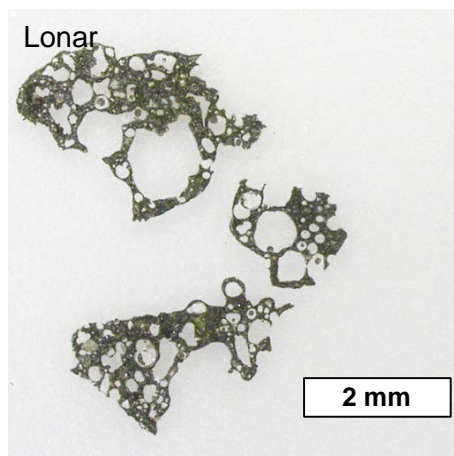

Figure 1

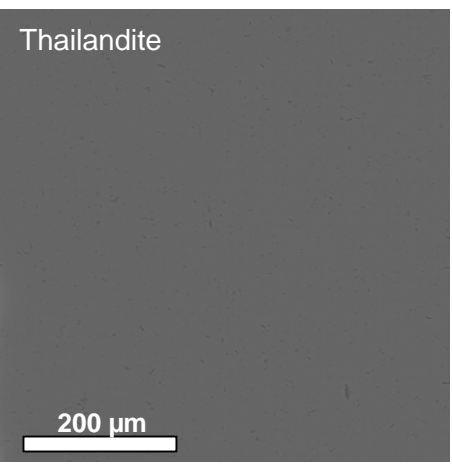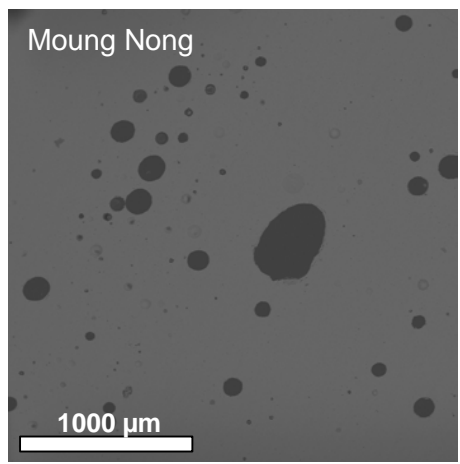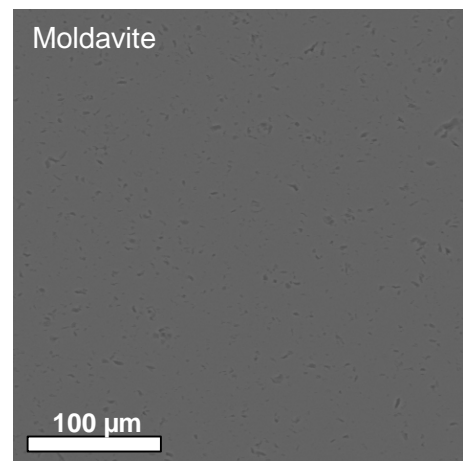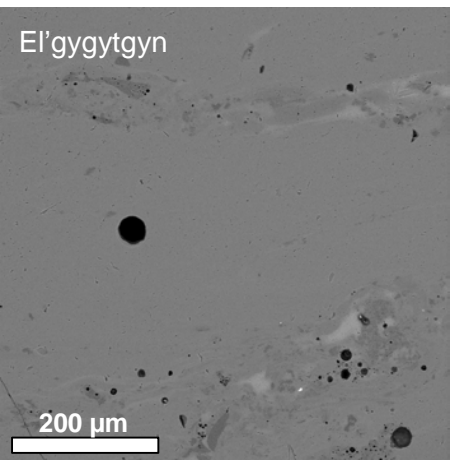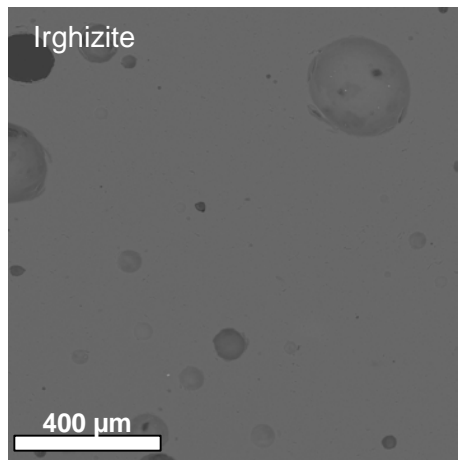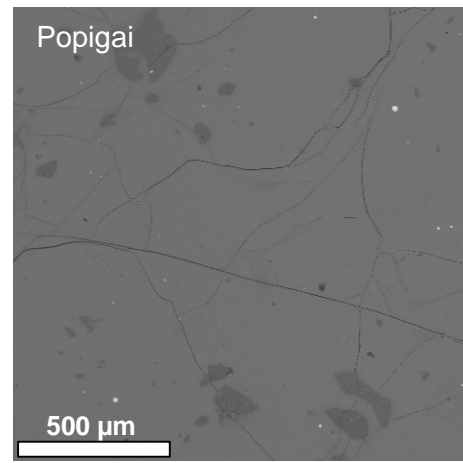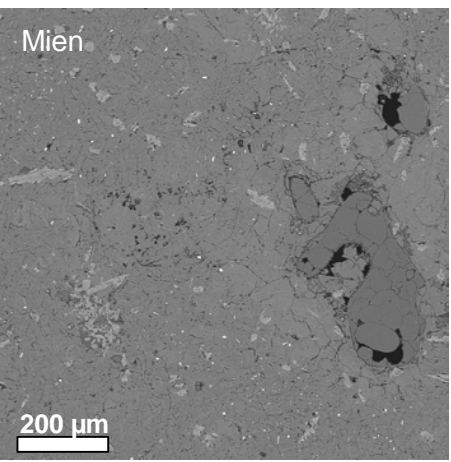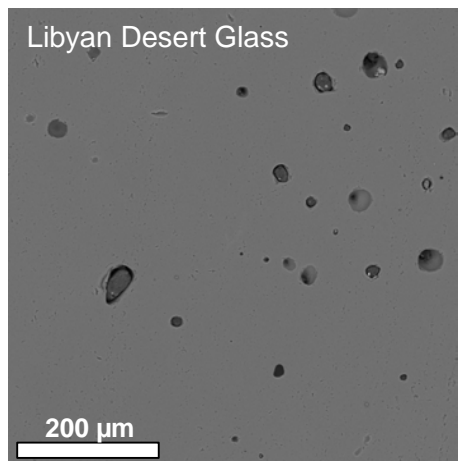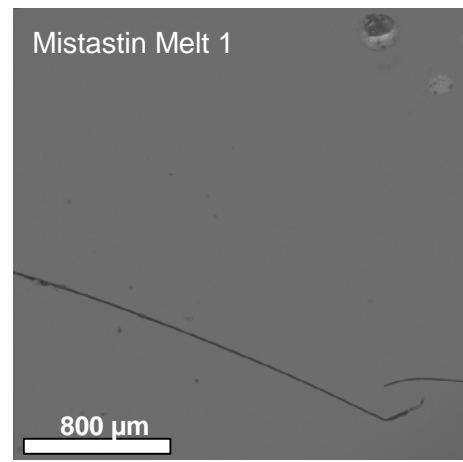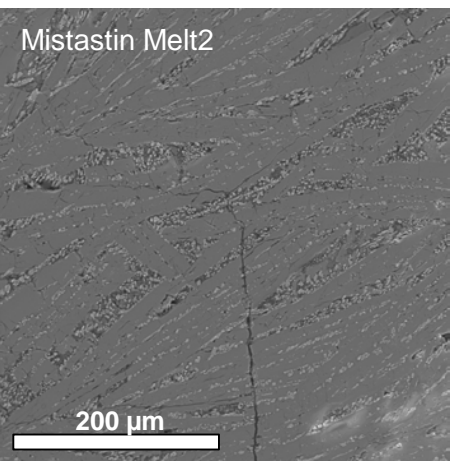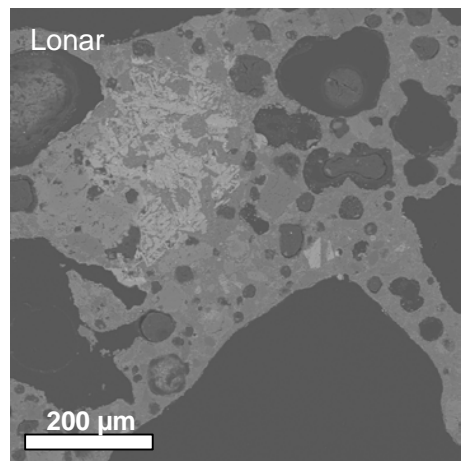

Figure 2

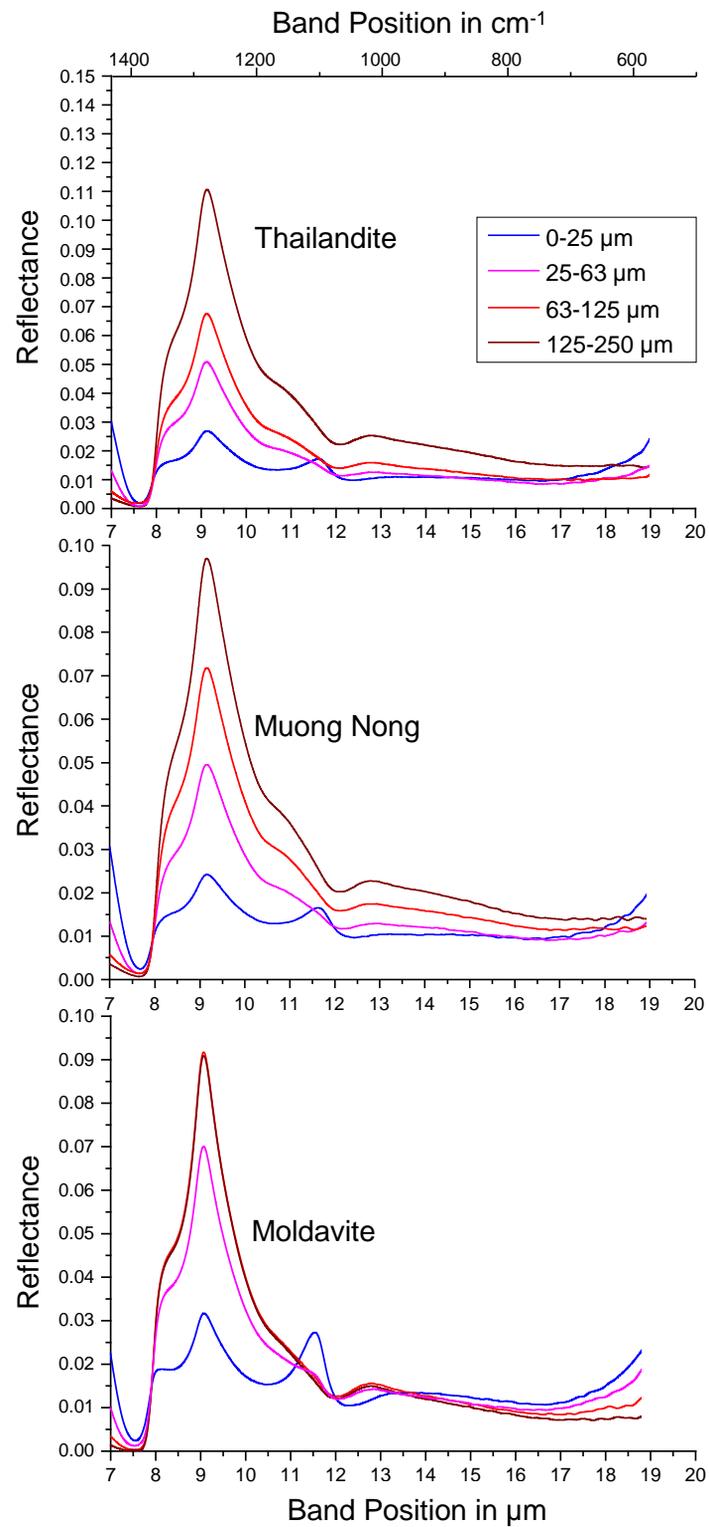

Figure 3

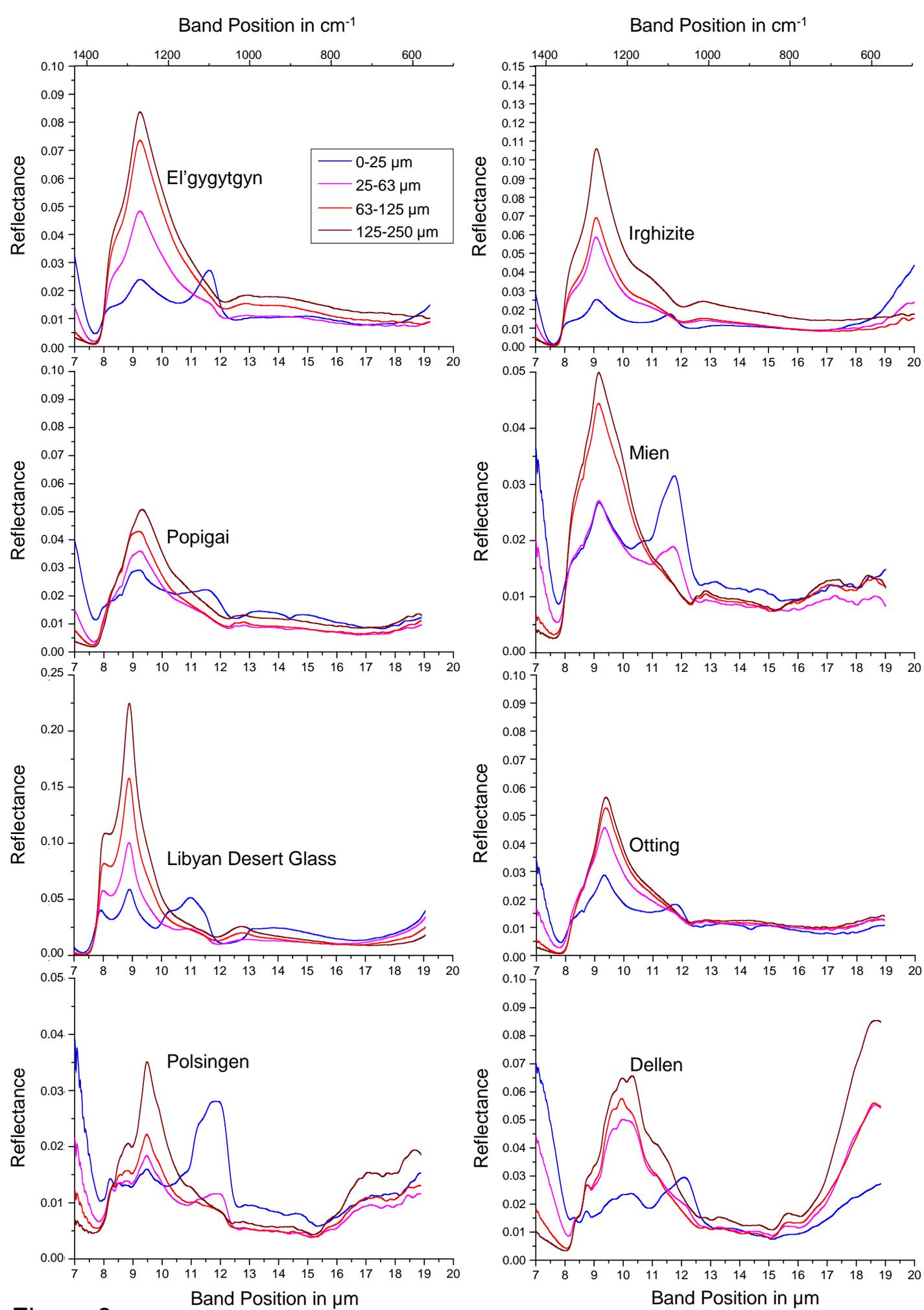

Figure 3

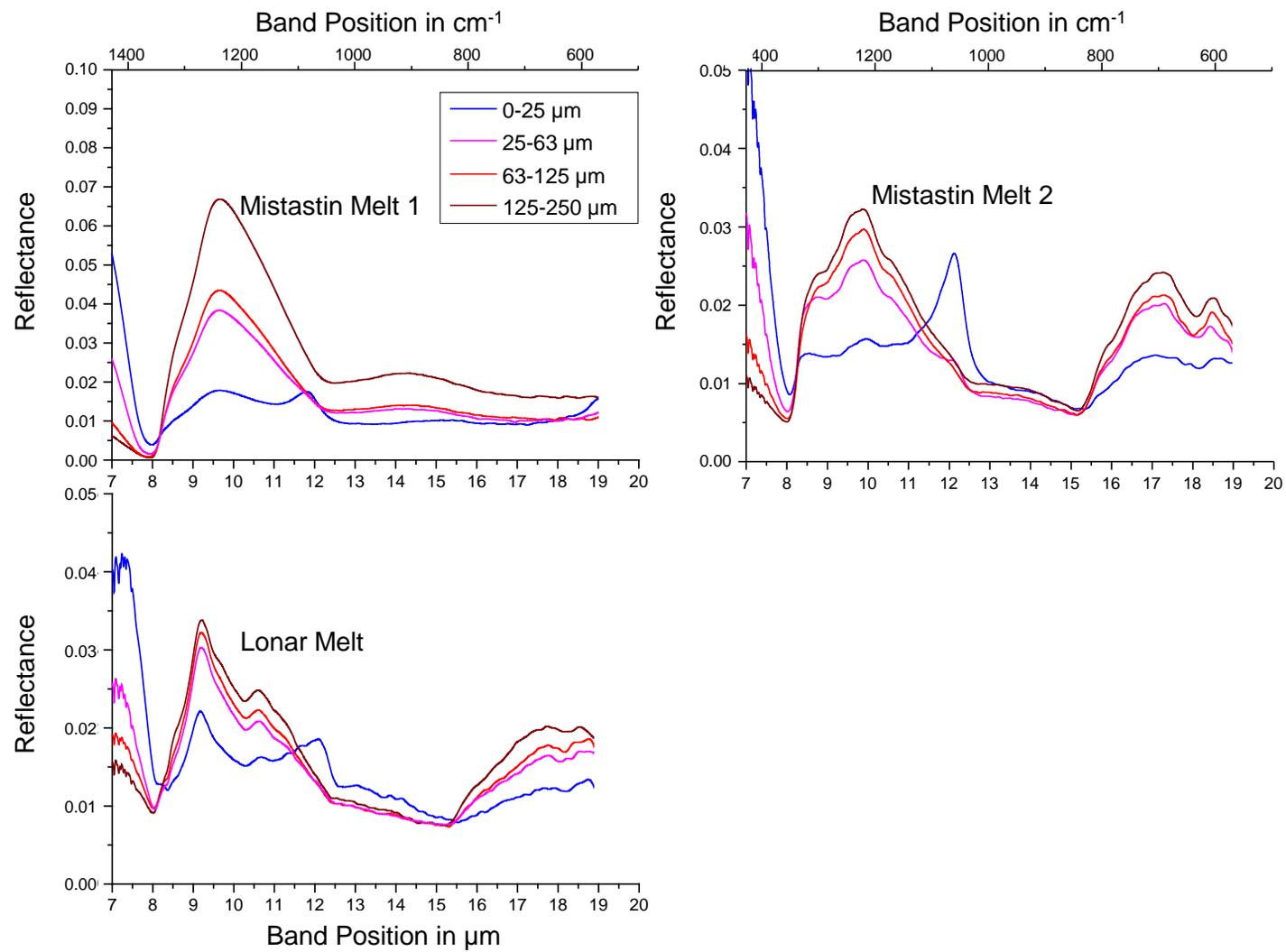

Figure 3

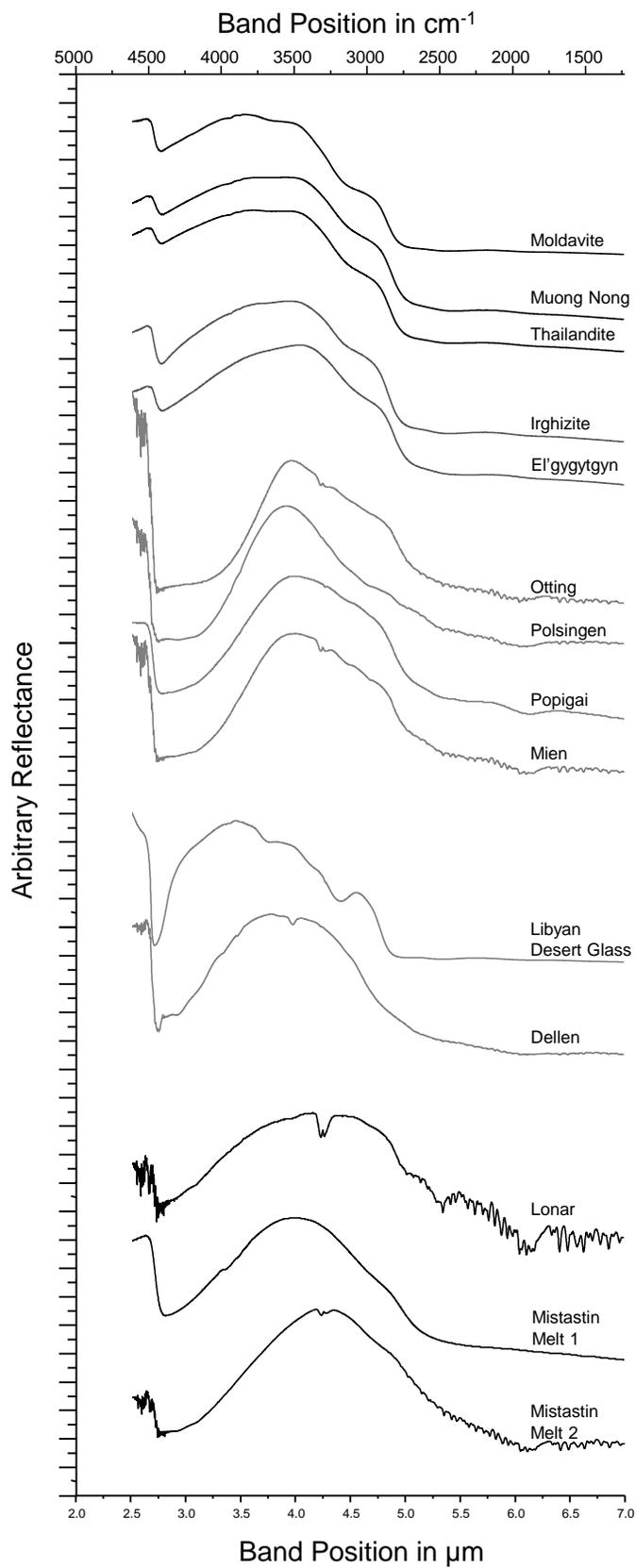

Figure 4

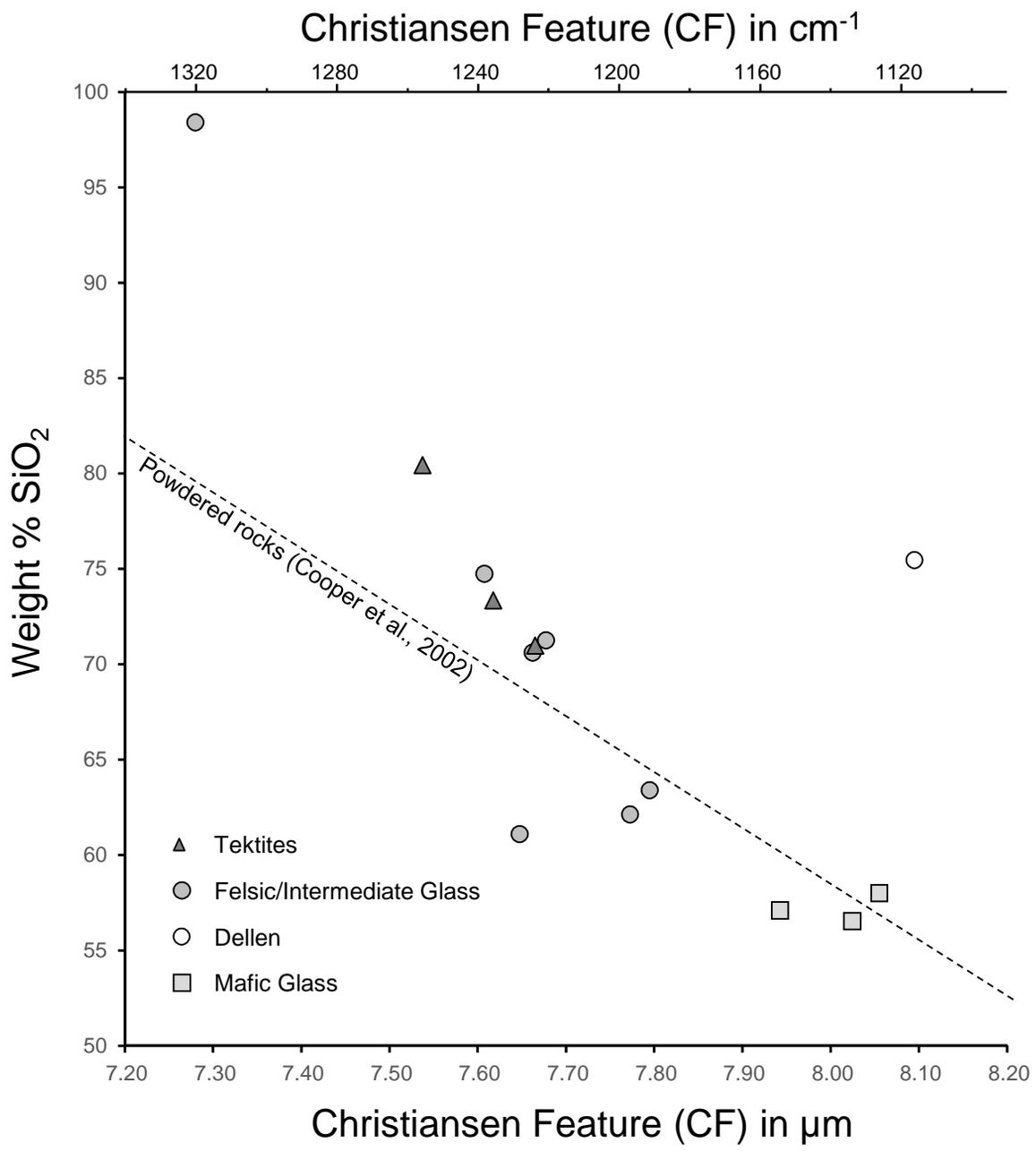

Figure 5

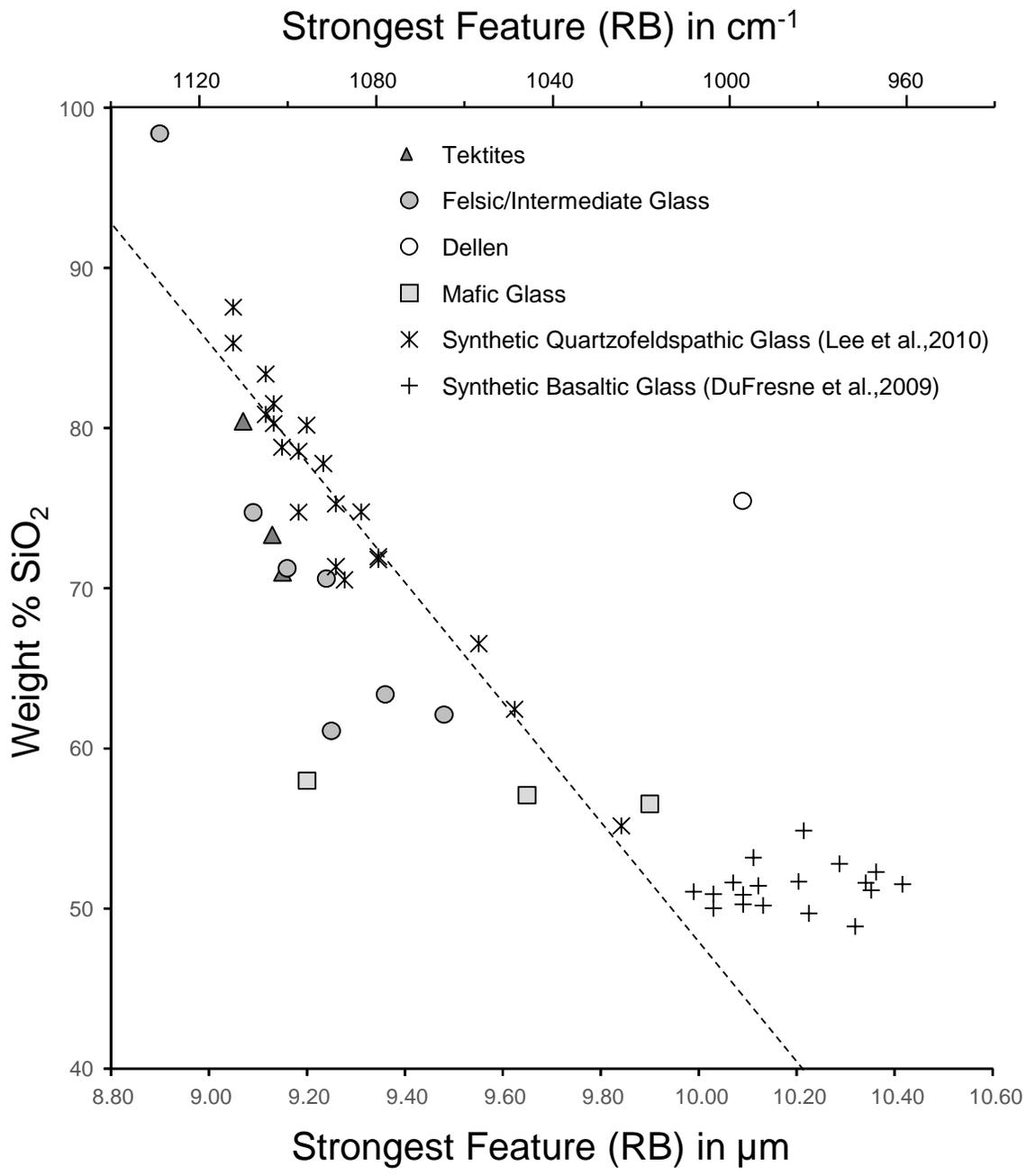

Figure 6

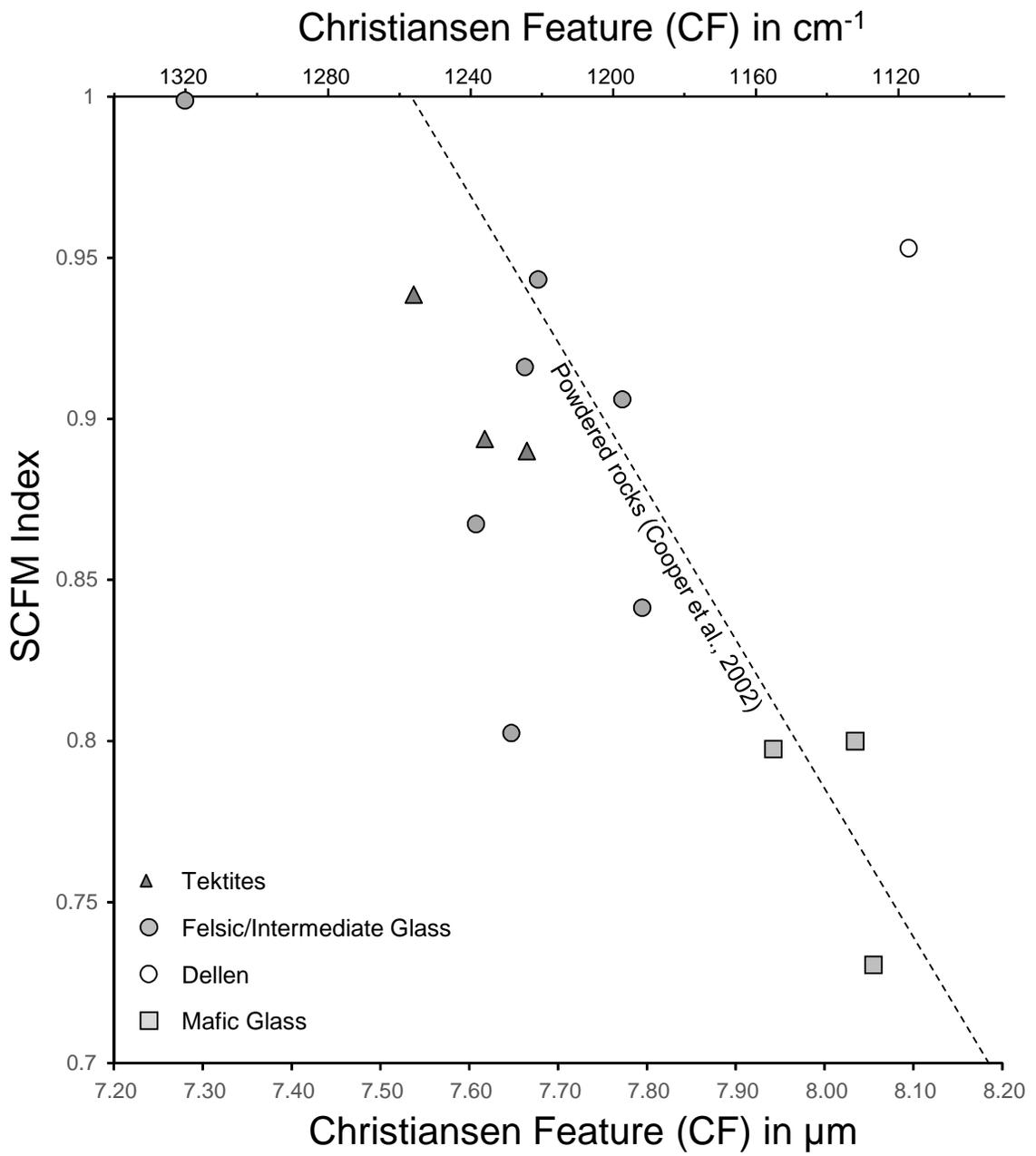

Figure 7